\documentstyle [preprint,aps]{revtex}
\begin{document}
\draft

\title
{Quantum-Classical Correspondence via Liouville Dynamics:
II. Correspondence for Chaotic Hamiltonian Systems}
\author{Joshua Wilkie and Paul Brumer}
\address{Chemical Physics Theory Group, Department of Chemistry,
   University of Toronto,
   Toronto, Ontario, Canada M5S 1A1 }

\maketitle

\begin{abstract}
	We prove quantum-classical correspondence for bound
conservative classically chaotic Hamiltonian systems.  In particular,
quantum Liouville spectral projection operators
and spectral densities, and hence classical dynamics, are shown to approach their classical analogs
in the $h\rightarrow 0$ limit.
Correspondence is shown to occur via the elimination of essential singularities.
In addition, applications to matrix elements of
observables in chaotic systems are discussed.

\end{abstract}

\section{Introduction}
\label{4.1}

The validity of quantum mechanics as a description of the macroscopic
world is contingent upon the reduction of the laws of quantum
mechanics to Newton's laws in the limit where the characteristic actions of a system are large with
respect to Planck's constant\cite{IVcp}. Thus, diagonal and off-diagonal
matrix elements must reduce to their classical analogs and quantum
dynamics must reproduce the predictions of classical mechanics as
$h\rightarrow 0$. Despite the fundamental importance of
quantum-classical correspondence it has only been satisfactorily
demonstrated\cite{IVberry1}-\cite{paper1} in the
very restrictive case of regular systems, i.e., systems
which classically possess as
many constants of the motion as degrees of freedom. Indeed some authors
have suggested that bound
quantum systems with discrete quantum spectrum and chaotic classical analog
may violate the correspondence principle\cite{IVford}. These doubts
about the validity of the correspondence principle for chaotic
systems stem from the difficulty of reconciling the quasiperiodic nature of bound state quantum
dynamics with the chaotic nature of classical dynamics for the same
Hamiltonian.  The issue of
correspondence for quantum systems whose classical analogs exhibit
chaos (irregular systems) is thus of great interest.

Verification of correspondence should be distinguished from the
development of semiclassical approximation methods. While semiclassical
theories provide a natural
starting point for an exploration of the classical limit their
existence does not guarantee correspondence. For example, semiclassical
theories for regular systems preceded the
development of modern quantum mechanics\cite{IVjam}, but an understanding of
correspondence for regular systems has only recently been
achieved\cite{IVberry1,IVjaffe2,paper1}.
By comparison, attempts to develop
semiclassical quantization rules
for chaotic systems have had some success\cite{IVclass}, whereas
the correspondence
limit remains largely unexplored\cite{IVberry4}. In this paper we
demonstrate that the existing semiclassical theories of
quantum dynamics for classically chaotic systems are sufficiently well 
developed to allow us to show that
such systems do in fact approach their proper correspondence
limits as Planck's constant approaches zero. This completes the
Liouville correspondence program outlined in the preceding 
paper\cite{paper1}, and significantly extends the results
of our study of quantum maps\cite{IVwbcat,patt}, where we rigorously demonstrated
that a nonchaotic
quantum map dynamics can completely recover a fully chaotic classical dynamics
in the limit $h\rightarrow 0$.

The Liouville picture affords a means of gaining insight into the
connections between quantum and classical
mechanics\cite{IVjaffe2,IVjaffe1,IVww}, and is a
natural framework for studies of correspondence. As outlined
in the preceding paper (henceforth referred to as Paper 1)
\cite{paper1} the essential
ingredients for Liouville dynamics are eigenstates and
eigenvalues of the Liouville operators in both mechanics. In
particular, the
dynamics is completely characterized by the Liouville eigenfunctions
and eigenvalues or the spectral projectors once the class of allowed
initial distributions is specified. Here we consider correspondence in
chaotic systems from this Liouville perspective.

Quantum Liouville eigenfunctions for conservative Hamiltonian systems
whose classical analogs are chaotic take the form $|n\rangle\langle m|$
where $|n\rangle$ are eigenstates of the Hamiltonian, i.e., $\hat{H}|n\rangle
=E_n|n\rangle$. These distributions are eigenfunctions of the complete
set of operators $\hat{L}, \hat{{\cal H}}$ where
$\hat{L}=\frac{1}{2}[\hat{H},]$ is the quantum Liouville operator and
where $\hat{{\cal H}}=\frac{1}{2}[\hat{H},~]_+$ is the Hermitian energy operator in the
Liouville picture\cite{paper1}.  That is, they are 
 solutions of both the time independent Liouville equation
 \begin{equation} 
\hat{L}|n\rangle\langle
 m|=\lambda_{n,m}|n\rangle\langle m|, 
\label{Leig}
\end{equation} where
 $\lambda_{n,m}=(E_n-E_m)/\hbar$, and of the energy
 eigenequation 
 \begin{equation} 
 \hat{{\cal H}}|n\rangle\langle
 m|=E_{n,m}|n\rangle\langle m|, 
\label{Heig}
 \end{equation} 
 with $E_{n,m}=(E_n+E_m)/2$.

Consistent with von Neumann's criteria for
quantum ergodicity \cite{IVvonN}, we
deal with quantum systems with chaotic classical analog\cite{IVhyper,IVarnold}
for which the spectrum of energies $E_n$ is nondegenerate. For
such systems the states $|n\rangle\langle m|$ are specified
by the integers $n$ and $m$, or equivalently by the frequency
$\lambda_{n,m}$ and energy $E_{n,m}$.
Since the
distributions $|n\rangle\langle m|$ govern the quantum
dynamics \cite{paper1} an understanding of their $h\rightarrow 0$ limit, or of their
Wigner representation $\rho^w_{n,m}({\bf x})$,
\begin{equation}
\rho_{n,m}^w({\bf x})\equiv h^{-s/2}\int d{\bf
v}e^{i{\bf p}\cdot{\bf v}/\hbar}\langle {\bf q}-{\bf
v}/2| n\rangle\langle m|{\bf q}+{\bf v}/2\rangle ,
\label{def1}
\end{equation}
would seem essential for verification of correspondence.
[Here ${\bf x}=({\bf p, q})$ where ${\bf p}$ are the momenta and 
${\bf q}$ are the coordinates.]
However, as shown below,
the relevant objects for the study of correspondence in chaotic
systems are the quantum spectral projection operators\cite{IVaphase} which are
 of the form 
$\rho_{n,m}^{w*}({\bf x}_0)\rho_{n,m}^w({\bf x})$ in the Wigner
representation. That is, we demonstrate that for irregular systems these
quantum Liouville spectral projection operators approach
classical spectral projection operators $\Upsilon$  of the same frequency and
energy as $h \rightarrow 0$, i.e., that
\begin{equation}
\rho_{n,n}^{w*}({\bf x}_0)\rho_{n,n}^w({\bf x})\rightarrow
dE~\Upsilon_{E_n}({\bf x},{\bf x}_0), 
\label{lim1}
\end{equation}
and 
\begin{equation}
\rho_{n,m}^{w*}({\bf x}_0)\rho_{n,m}^w({\bf x})\rightarrow
dEd\lambda~\Upsilon_{E_{n,m},\lambda_{n,m}}({\bf x};{\bf x}_0),  ~~(n\neq m).
\label{lim2}
\end{equation}
Here the distributions $\Upsilon_E$ and
$\Upsilon_{E,\lambda}$ are the stationary and nonstationary Liouville
spectral projection operators of classical dynamics\cite{paper1},
discussed in Paper 1.  We also show that the spectrum of the
quantum Liouville operator goes to that of the classical operator as
$h \rightarrow 0$ and that the correspondence emerges smoothly
via the elimination of essential singularities.
 Proof of Eqs. (\ref{lim1}) and (\ref{lim2}), plus proof
of the correspondence of the Liouville spectra, suffices to prove
quantum-classical correspondence in chaotic systems.
 
 Note that,
unlike their quantum analogs $\rho_{n,m}^{w*}({\bf x}_0)\rho_{n,m}^w({\bf x})
$, the nonstationary chaotic classical spectral
projection operators $\Upsilon_{E_{n,m},\lambda_{n,m}}$ cannot be written as a product
of Liouville eigenfunctions but rather consist of a 
sum of products of Liouville eigenfunctions (see Paper 1). Hence Eq.
(\ref{lim2}) suggests, as discussed below, that
the individual nonstationary quantum Liouville eigenfunctions
$\rho^w_{n,m}({\bf x})$ for
quantum systems with chaotic classical analogs do not have well defined
correspondence limits. This situation is quite different from that of
the integrable case discussed in Paper 1 and necessitates the introduction
of new tools to prove correspondence. 

This paper is organized as follows: Section \ref{4.2} introduces a
useful Dirac notation to simplify our formal manipulations, and
the proof of Eqs. (\ref{lim1}) and (\ref{lim2})
is expressed in this Dirac form. Section \ref{4.4} proves correspondence for both the Liouville spectral projectors
and the Liouville eigenvalues. This treatment ignores higher order
corrections relating to scars, which are treated in Section \ref{4.5}.
The proof of correspondence allows us to consider the classical limit of 
operator matrix elements, which is discussed in Section \ref{4.6}.
Section \ref{4.7} provides a summary.
 
\section{A Dirac Formulation of Liouville Dynamics}
\label{4.2}

The effectiveness of the Liouville picture is limited by the
clumsiness of the associated density matrix notation.
 In this section we introduce a useful Dirac notation which simplifies
 manipulations considerably \cite{IVdform}. We will
also employ a Dirac
notation for the classical Liouville dynamics in order to
maintain symmetry between the quantum and
classical formulations.

Let $|n\rangle$ be a
complete, orthonormal set of basis states for the quantum
Hilbert space associated with the solutions of the Schr\"{o}dinger
equation. That is, $\langle n|m\rangle=\delta_{n,m}$ and
$\sum_{n}|n\rangle\langle n|=1$. From these states we construct
distributions
$\hat{\rho}_{n,m}=|n\rangle\langle m|$ which are a basis in
the Hilbert space associated with solutions of the von Neumann
equation. It is natural to assign a Dirac notation to these
basis states, i.e.,
\begin{equation}
|n,m
)\equiv \hat{\rho}_{n,m}. 
\label{dket}
\end{equation}
A complete
orthonormal basis $|n\rangle$ of Schr\"{o}dinger states then yields a
complete set of Liouville states $|n,m)$. One can now easily deduce that
the dual space is spanned by the linear functionals
\begin{equation}
(n,m|={\rm Tr}\{\hat{\rho}_{n,m}^{\dag}\cdot\}
\label{dbra}
\end{equation}
by requiring
that $(n,m|k,l)=\delta_{n,m}\delta_{k,l}$. Note that the normalization
of the states $|n,m)$ has been chosen so that $|n,m)(n,m|$ is a
projection operator. Completeness
implies that 
\begin{equation}
\sum_{n,m}|n,m)(n,m|=1.
\end{equation}
The spectral
decomposition of
the Liouville operator then takes the form
\begin{equation}
\hat{L}=\sum_{n,m}\lambda_{n,m}|n,m)(n,m|.
\label{ldecomp}
\end{equation}

The Hermitian operators\cite{IVherm} $|n,m)(n,m|=\hat{\rho}_{n,m}{\rm
Tr}\{\hat{\rho}_{n,m}^{\dag}\cdot\}$ are obviously the spectral projection
operators in the Liouville picture, in the same way that
$|n\rangle\langle n|$ are the spectral projection operators in the
traditional
Hamiltonian picture. Arbitrary superoperators of the
form $[\hat{O},~]_{\pm}$ (i.e.,
$[\hat{O},~]_{\pm}\hat{\rho}=\hat{O}\hat{\rho}\pm \hat{\rho}\hat{O}$),
of interest below,
can be expanded on the $|n,m)$ states as
\begin{equation}
[\hat{O},~]_{\pm}=\sum_{n,m,k,l}|n,m)(n,m|[\hat{O},~]_{\pm}|k,l)(k,l|
\end{equation}
where the superoperator ``matrix elements'' are
$(n,m|[\hat{O},~]_{\pm}|k,l)=O_{n,k}\delta_{l,m}\pm O_{l,m}\delta_{n,k}$.

Physical states
$|\rho)$ are defined as
\begin{equation}
|\rho)=h^{-s/2}\hat{\rho}, 
\label{pket}
\end{equation}
with
corresponding kets 
\begin{equation}
(\rho|=h^{-s/2}{\rm Tr}\{\hat{\rho}^{\dag}\cdot\}=h^{-s/2}{\rm
Tr}\{\hat{\rho}\cdot\},  
\label{pbra}
\end{equation}
with the latter equality due to the fact that $\hat{\rho}^{\dag}=\hat{\rho}$.
Equations (\ref{pket}) and (\ref{pbra}), which define the physical
states, differ from Eqs. (\ref{dket}) and (\ref{dbra}) which define the
basis states by a factor of $h^{-s/2}$ which is introduced so
that the quasi-probability distributions associated with the physical
states have the correct dimensions in a phase space representation,
i.e., inverse action to a power
equal to the number of degrees of freedom.
We also assign states $|A)$ to operators $\hat{A}$ (i.e., operators
operating on the Hilbert space spanned by $ |n \rangle)$ via
$|A)=h^{s/2}\hat{A}$ and $(A|=h^{s/2}{\rm Tr}\{\hat{A}^{\dag}\cdot\}$. The
expectation of $\hat{A}$ is then given by $(\rho|A)=(A|\rho)^*={\rm
Tr}\{\hat{\rho}\hat{A}\} $.

In this notation the von Neumann (quantum Liouville)
equation\cite{IVobs} is
\begin{equation}
\frac{\partial}{\partial t}|\rho (t))=-i\hat{L}|\rho (t)). 
\end{equation}
and the Wigner-Weyl representation\cite{IVdirec,IVbj} of a state $\rho$ takes the form 
\begin{equation}
({\bf x}|\rho)=h^{-s/2}{\rm
Tr}\{h^{-s/2}\hat{\Delta}({\bf x})\hat{\rho}\},
\end{equation} 
where 
\begin{equation}\hat{\Delta}({\bf
x})=h^{-s}\int d{\bf u}d{\bf v}~e^{i[{\bf v}\cdot ({\bf p}-\hat{{\bf
p}})+{\bf u}\cdot ({\bf q}-\hat{{\bf q}})]/\hbar}.
\end{equation}
Thus, employing Eq. (\ref{pket}), we identify 
\begin{equation}
({\bf x}|={\rm Tr}\{h^{-s/2}\hat{\Delta}({\bf x})\cdot \}=h^{-s/2}{\rm Tr}
\{\hat{\Delta}^{\dag}({\bf x})\cdot \}
\label{xbra}
\end{equation}
where the second equality is due to the fact that $\hat{\Delta}({\bf x})$ is 
Hermitian. The
particular form of the corresponding ket \cite{IVrl,IVmiz} is determined by
demanding that $({\bf x}|{\bf x}')=\delta({\bf x}-{\bf x}')$. Thus,
here
\begin{equation}
|{\bf x})=h^{-s/2}\hat{\Delta}({\bf x}), 
\label{xket}
\end{equation}
where $({\bf x}|{\bf x}')= h^{-s}{\rm Tr}\{{\hat{\Delta}}({\bf
x}){\hat{\Delta}}({\bf x}') \}=\delta({\bf x}-{\bf x}')$. 
Since $|{\bf x})$ and $({\bf x}|$ span 
the Hilbert space and its dual space, they satisfy the closure relation
\begin{equation}
\int d{\bf x}~ |{\bf x})({\bf x}|=1.
\label{xclosure}
\end{equation}
Definitions
(\ref{xbra}) and (\ref{xket}) in conjunction with Eqs.
(\ref{pket}) and (\ref{pbra}) 
guarantee that the probability densities $({\bf x}|\rho
)$ have the correct dimensions.
Other phase space representations [in which $({\bf x}|$ and $|{\bf
x})$ may be quite dissimilar], and a general transformation theory
between them is provided elsewhere\cite{IVwbreps}.

Consider then the Liouville spectral decomposition [i.e., Eqs.
(\ref{Leig}) and (\ref{Heig})] for a chaotic quantum system in the
Dirac notation. As eigenfunctions of $\hat{L}$ and $\hat{\cal H}$ the
$|n,m)$ satisfy
\begin{equation}
\hat{L}|n,m)=\lambda_{n,m}|n,m),
\label{eqq6}
\end{equation}
and
\begin{equation}
\hat{{\cal H}}|n,m)=E_{n,m}|n,m).
\label{wweig2}
\end{equation}
 In the Wigner-Weyl
representation Eqs. (\ref{eqq6}) and (\ref{wweig2}) become 
\begin{equation}
({\bf x}|\hat{L}|n,m)=L({\bf x})({\bf
x}|n,m)=\lambda_{n,m}({\bf x}|n,m),
\end{equation}
and
\begin{equation}
({\bf x}|\hat{{\cal H}}|n,m)={\cal H}({\bf x})({\bf
x}|n,m)=E_{n,m}({\bf x}|n,m),
\end{equation}
where $L({\bf x})=\frac{2i}{\hbar}H({\bf
x})\sin (\hbar\sigma/2)$ is the quantum Liouville operator, 
and ${\cal H}({\bf x})=H({\bf x})\cos \{\hbar \sigma/2\}$ is the 
energy operator. Here $\sigma=\widetilde{\frac{\partial^{\leftarrow}}{\partial
{\bf x}}}\cdot J\cdot \frac{\partial^{\rightarrow}}{\partial {\bf x}}$ is the
Poisson bracket, i.e., $A({\bf x})\sigma B({\bf x})=\{A,B\}$, and $J=\left( \begin{array}{cc}
0&-I \\
I &0
\end{array} \right)$ is the $2s\times 2s$ dimensional symplectic
matrix\cite{IVgold}. Expanding ${\cal H}({\bf x})$ in powers of $h$ shows that the classical
analog of ${\cal H}({\bf x})$ is the energy function $H({\bf x})$, and
that  
$L_c({\bf x})$ is the correspondence limit of $L({\bf x})$.

Consider now the classical case. The classical analog of a phase space
representation is a choice of canonical
variables for a classical distribution $\rho_c$. Thus we denote the phase space
representation
of $\rho_c$ by  $\rho_c({\bf x})=({\bf x}|\rho_c)$. 

The classical Liouville spectral decomposition, and the properties of
the eigendistributions discussed in Paper 1 \cite{paper1} are readily restated using the Dirac notation. Associating states $|E)$ with the classical distributions
$\rho_E({\bf x})$ which span the point spectrum, and
states $|E,\lambda,\ell)$ with the classical distributions
$\rho_{E,\lambda}^{\ell}({\bf x})$ which span the continuous spectrum, the
full set of equations for the spectral decomposition becomes:
\begin{equation}
(E'|E)=\delta (E'-E),
\label{eqc1}
\end{equation}
\begin{equation}
(E'|E,\lambda,\ell)=0,
\end{equation}
\begin{equation}
(E'\lambda',\ell'|E,\lambda,\ell)=
\delta_{\ell',\ell}\delta(E'-E)\delta(\lambda'-\lambda),
\end{equation}
\begin{equation}
\int_0^{\infty}dE
~|E)(E|+\int_0^{\infty}dE\int\!\!\!\!\!\!- d\lambda \sum_{\ell}
|E,\lambda,\ell)(E,\lambda,\ell|=1,
\label{cclos}
\end{equation}		 
\begin{equation}
e^{-iL_ct}|E)=|E),
\end{equation}
and
\begin{equation}
e^{-iL_ct}|E,\lambda,\ell)=e^{-i\lambda t}|E,\lambda,\ell).
\label{eqc6}
\end{equation}
Here the line through the integral in Eq. (\ref{cclos}) indicates that the 
point spectrum eigenvalue $\lambda=0$ has been removed (see Paper 1).
Two further equations relate to the second constant of the motion, a classical
energy operator ${\cal H}_c$:
\begin{equation}
{\cal H}_c|E)=E|E),
\label{bore}
\end{equation}
and
\begin{equation}
{\cal H}_c|E,\lambda,\ell)=E|E,\lambda,\ell).
\label{slop}
\end{equation}
In the phase space representation parameterized by ${\bf x}$ these
equations become
\begin{equation}
({\bf x}|{\cal H}_c|E)=H({\bf x})({\bf
x}|E)=E({\bf x}|E),
\label{cleq1}
\end{equation}
and
\begin{equation}
({\bf x}|{\cal H}_c|E,\lambda,\ell)=H({\bf x})({\bf
x}|E,\lambda,\ell)=E({\bf x}|E,\lambda,\ell).
\label{cleq2}
\end{equation}

A complete set of stationary and nonstationary classical Liouville
eigenfunctions $\rho_E({\bf x})=({\bf x}|E)$ and 
 $\rho_{E,\lambda}^\ell ({\bf x}) =
({\bf x}|E,\lambda, \ell)$, were introduced in Paper 1 
where the integer $\ell$ labels the infinite degeneracy of the
continuous spectrum\cite{IVarnold}. In addition, spectral projection
operators $\Upsilon_E({\bf x};{\bf x}_0)$ and 
$\Upsilon_{E,\lambda}({\bf x};{\bf x}_0)$ were introduced;
these are the phase space representations of the classical
operators $\delta(E-{\cal H}_c)=|E)(E|$
and $\delta(E-{\cal H}_c) \delta(\lambda-L_c)=\sum_{\ell}|E,\lambda ,\ell)(E,\lambda ,\ell|$. Specifically,
\begin{equation}
({\bf x}|E)(E|{\bf x}_0) =
\Upsilon_E({\bf x};{\bf x}_0)=
\frac{\delta(E-H({\bf x}_0))\delta(E-H({\bf x}))}{\int d{\bf x}'\delta
(E-H({\bf x}'))} =\rho_E({\bf x}) \rho_E^*({\bf x}_0)
\end{equation}
and 
\begin{eqnarray}
& & \sum_{\ell}~
({\bf x}|E,\lambda,\ell)(E, \lambda,\ell|{\bf x}_0)=
\Upsilon_{E,\lambda}({\bf x};{\bf x}_0)\nonumber \\
& &=\frac{1}{2\pi}\delta(E-H({\bf
x}_0))\int_{-\infty}^{\infty}dt^{\prime}e^{i\lambda
t^{\prime}}\delta({\bf x}_0-{\bf X}({\bf x},-t^{\prime}))
=\sum_{\ell}~ \rho^{\ell *}_{E,\lambda} ({\bf x}_0)
\rho^{\ell}_{E,\lambda} ({\bf x}),
\label{eq26}
\end{eqnarray}
where ${\bf X}({\bf x},-t^{\prime})$ is the phase space point from
which ${\bf x}$ emerges over a time $t'$.
 
 In terms of these eigenfunctions the spectral decomposition [Eq.
(\ref{cclos})] takes the
form 
\begin{equation}
\int_0^{\infty} dE~\rho_E^*({\bf
x}_0)\rho_E({\bf x})+\int_0^{\infty} dE\int\!\!\!\!\!\!- d\lambda
~\sum_{\ell}\rho_{E,\lambda}^{\ell *}({\bf
x}_0)\rho_{E,\lambda}^{\ell} ({\bf x})
=\delta({\bf x}-{\bf x}_0),
\end{equation}
or
\begin{equation}
\int_{0}^{\infty}dE ~\Upsilon_E ({\bf x};{\bf x}_0)+
\int_{0}^{\infty} dE\int\!\!\!\!\!\!- d\lambda ~\Upsilon_{E,\lambda}
({\bf x};{\bf x}_0)=\delta({\bf x}-{\bf x}_0).
\end{equation}

Thus, the evolution of any initial distribution $\rho({\bf x},0)$ can be
written, in quantum mechanics, as an expansion:
\begin{eqnarray}
\rho({\bf x},t) &=& \sum_n c_{n,n} \: \rho_{n,n}^w ({\bf x})+\sum_{n\neq m}
 c_{n,m} \: \rho_{n,m}^w ({\bf x})
e^{-i\lambda_{n,m} t} \nonumber \\
&=& \int d {\bf x}_0 \rho ({\bf x}_0,0) \: [\sum_n
\rho_{n,n}^{w*}({\bf x}_0) \rho^w_{n,n}({\bf x})
+\sum_{n \neq m}\rho_{n,m}^{w*}({\bf x}_0) \rho^w_{n,m}({\bf x})
 e^{-i\lambda_{n,m} t}]
\label{eq:5}
\end{eqnarray}
and in classical mechanics as
\begin{eqnarray}
\rho^c({\bf x},t) &=& \int dE ~c_E~\rho_E({\bf x})+\int 
d E \int\!\!\!\!\!\!-
 d \lambda \sum_{\ell} c_{E,\lambda,\ell} 
\: \rho_{E,\lambda}^{\ell} ({\bf x})
e^{-i\lambda t} \nonumber \\
&=& \int d {\bf x}_0 \rho({\bf x}_0,0) [\int dE~\rho^{*}_E({\bf
x}_0)\rho_E({\bf x})+\int d E\int\!\!\!\!\!\!- d\lambda\sum_{\ell} 
\rho_{E,\lambda}^{\ell*}({\bf x}_0)
\rho_{E,\lambda}^{\ell}({\bf x}) e^{-i\lambda t}]\nonumber \\ 
&=& \int d {\bf x}_0 \rho({\bf x}_0,0) [\int dE~\Upsilon_E({\bf x};{\bf
x}_0)+\int d E\int\!\!\!\!\!\!- d\lambda \Upsilon_{E,\lambda}({\bf x};{\bf
x}_0)  e^{-i\lambda t}].
\label{eq:6}
\end{eqnarray}
Equations (\ref{eq:5}) and (\ref{eq:6}) make clear that a
demonstration of correspondence for the spectral projection operators
and their eigenvalues is sufficient to establish correspondence for the
dynamics, i.e., that $\rho({\bf x},t)
\rightarrow \rho^c({\bf x},t) $ as $h\rightarrow 0$.
That is, formally establishing correspondence requires demonstrating
\begin{equation}
|n,n)(n,n|\rightarrow dE ~|E_n)(E_n|
\label{corr1}
\end{equation}
and
\begin{equation}
|n,m)(n,m|\rightarrow dEd\lambda ~\sum_{\ell}
~|E_{n,m},\lambda_{n,m},\ell)(E_{n,m},\lambda_{n,m},\ell|,
\label{corr2}
\end{equation}
or
\begin{equation}
({\bf x}|n,n)(n,n|{\bf x}_0)=\rho_{n,n}^{w*}({\bf x}_0) \rho^w_{n,n}({\bf x})
\rightarrow dE~\Upsilon_{E_n}({\bf x};{\bf
x}_0)=dE~({\bf x}|E_n)(E_n|{\bf x}_0)
\label{statc}
\end{equation}
and
\begin{eqnarray}
({\bf x}|n,m)(n,m|{\bf x}_0)=
\rho_{n,m}^{w*}({\bf x}_0) \rho^w_{n,m}({\bf x})\rightarrow &&dEd\lambda
~\Upsilon_{E_{n,m},\lambda_{n,m}}({\bf x};{\bf x}_0)\nonumber \\
&&=dEd\lambda~\sum_{\ell}~({\bf
x}|E_{n,m},\lambda_{n,m},\ell)(E_{n,m},\lambda_{n,m},\ell|{\bf x}_0),
\label{nm}
\end{eqnarray}
with the infinitesimals $dE$ and $d\lambda$ to be
determined. These limits are proven in Sec. \ref{4.4.1}.

\section{Correspondence}
\label{4.4}

Consider then the correspondence limit, i.e., the limit of the quantum
Liouville dynamics as $h\rightarrow 0$, 
with the $h\rightarrow 0$ limit taken before the $T\rightarrow \infty$
limit \cite{IVberry4,IVrb}. This order, $h\rightarrow
0$ first, is
consistent with the actual physics in which one first chooses a
particular system and then propagates it for long times.   
Technically, this is achieved by first broadening the system energy by some
amount $\epsilon \gg h/T_{{\rm min}}$ (thus restricting the dynamics to
finite time) and taking the $h\rightarrow 0$ limit with $\epsilon$ fixed.
The broadening $\epsilon$ can then be chosen infinitesimal, 
$\epsilon \rightarrow \langle \Delta E \rangle$,
(where $\langle \Delta E \rangle$ is the average spacing between neighboring
energy levels) allowing for long time dynamics.
Here $T_{\rm min}$ is the period of the shortest periodic orbit.

The physical significance of
correspondence under these limits is clear. A transition from quantum
to classical behavior will be observed in the dynamics of a physical
system as $h\rightarrow 0$ provided that (a) the
apparatus with which we observe its dynamics has a
fixed, classically small but quantum mechanically large, energy
resolution, and that (b) we do not observe its dynamics beyond the
recurrence time given approximately by $h/\langle\Delta E\rangle$.

\subsection{Correspondence for Spectral Projection Operators}
\label{4.4.1}

Here we examine the correspondence limits of the
spectral projection operators $|n,n)(n,n|$ and
$|n,m)(n,m|$. 
We focus attention on the nonstationary
case. The stationary
 case [Eq. (\ref{statc})] has already been obtained by Berry and
Voros\cite{IVberry2,IVvoros,IVberry3}, but we work through this case
 to demonstrate the consistency of our
approach.  In the latter case consider Berry's formula \cite{IVberry2,IVberry3,IVformula}
\begin{equation}
\lim_{\epsilon\rightarrow 0}\pi\epsilon W({\bf x};E_n,\epsilon)
=\int d{\bf q}'e^{i{\bf p}\cdot{\bf q}'/\hbar}\langle {\bf
q}-{\bf q}'/2|n\rangle\langle n|{\bf q}+{\bf q}'/2\rangle
=h^{s/2} ~({\bf x}|n,n)
\label{subid1p}
\end{equation}
for finite $\epsilon$, in order to investigate the semiclassical
form of the stationary Liouville eigendistribution $({\bf x}|n,n)$
in the Wigner representation. Here $W({\bf x};E,\epsilon)$ is the Lorentzian
weighted sum of Wigner functions over a width
$\epsilon$ about an energy $E$, i.e., 
\begin{equation}
W({\bf x};E,\epsilon)\equiv \int d{\bf q}'e^{-i{\bf p}\cdot{\bf
q}'/\hbar}\langle {\bf q}+{\bf q}'/2|\delta_{\epsilon}(E-\hat{H})|{\bf
q}-{\bf q}'/2\rangle,
\end{equation}
where
\begin{equation}
\delta_{\epsilon}(E-\hat{H})=-\frac{1}{\pi}{\rm Im}\frac{1}{E-\hat{H}+i\epsilon}=\frac{1}{h}\int_{-\infty}^{\infty}dt
~e^{i(E-\hat{H})t/\hbar}e^{-\epsilon |t|/\hbar}.
\label{ids}
\end{equation}
For quantum systems with chaotic classical analogs,
Berry\cite{IVberry3} has shown, for small $h$ and small $\epsilon$,
that (where $\sim$ denotes the form in the limit)
\begin{equation}
W({\bf x};E,\epsilon)\sim\delta_{\epsilon}(E-H({\bf x}))+\sum_jW_{{\rm
scar}}^j({\bf x};E,\epsilon),
\label{bf}
\end{equation}
and thus, by employing Eq. (\ref{subid1p}), that
\begin{equation}
({\bf x}|n,n)\sim \pi\epsilon h^{-s/2}\delta_{\epsilon}(E_n-H({\bf x}))
+\pi\epsilon h^{-s/2}\sum_jW_{{\rm
scar}}^j({\bf x};E_n,\epsilon).
\label{eigb}
\end{equation}
Here $W_{{\rm scar}}^j$ is of the order $h^{s-1}$ smaller than the $\delta_{\epsilon}(E_n-H({\bf x}))$ term, and hence vanishes
rapidly as $h\rightarrow 0$. These scar terms, neglected in this
section, are considered in Sec. \ref{4.5}. 

Neglecting the scar terms gives
\begin{equation}
({\bf
x}|n,n)\sim h^{-s/2}\pi\epsilon \delta(E_n-H({\bf x}))\sim
\pi\epsilon ({\bf x}|E_n)/ \langle \Delta E\rangle^{1/2},
\end{equation}
where
$({\bf x}|E_n)=\rho_{E_n}({\bf x})=\delta(E_n-H({\bf x}))/[\int d{\bf
x}'\delta(E_n-H({\bf x}'))]^{1/2}$ (see Paper 1) and where
 $\langle \Delta E\rangle=h^s/\int d{\bf x}'\delta 
(E-H({\bf x}'))$ is the
average adjacent energy level spacing.  Therefore
\begin{equation}
({\bf x}|n,n)(n,n|{\bf x}_0)\sim
(\pi\epsilon)^2({\bf x}|E_n)(E_n|{\bf x}_0)/\langle \Delta E\rangle.
\label{statlim}
\end{equation}
Correspondence for the stationary eigenstates [Eq. (\ref{statc})] 
then results if we take the limit
$\pi\epsilon\rightarrow \langle \Delta E\rangle$ and note that
$\langle\Delta E\rangle \rightarrow dE$.

The proof of Eq. (\ref{nm}) follows in a similar fashion from the
following important relationship (proven in Appendix
A) between stationary and nonstationary quantum Liouville eigenfunctions 
\cite{IVwform}:
\begin{eqnarray}
({\bf x}|n,m)(n,m|{\bf x}_0)&=&h^{-s}\int d{\bf u}d{\bf v}e^{-i({\bf p}-{\bf p}_0)\cdot{\bf
v}/\hbar}e^{i({\bf q}-{\bf q}_0)\cdot{\bf u}/\hbar}\cdot\nonumber \\
&&(({\bf
p}+{\bf u}/2,{\bf q}+{\bf v}/2)|n,n)(m,m|({\bf p}_0-{\bf u}/2,{\bf
q}_0-{\bf v}/2))\nonumber \\
&=&h^{-s}\int d{\bf y}e^{i({\bf x}-{\bf x}_0)\cdot J\cdot\tilde{{\bf y}}/
\hbar}\cdot ({\bf x}+{\bf y}/2|n,n)(m,m|{\bf x}_0-{\bf y}/2).
\label{subid2p}
\end{eqnarray}
Here the tilde denotes the transpose (a column vector) of the row vector
${\bf y}$. Substituting
Eq.~(\ref{subid1p}) into Eq.~(\ref{subid2p}) gives:
\begin{eqnarray}
&&({\bf x}|n,m)(n,m|{\bf x}_0)
=\lim_{\epsilon_1,\epsilon_2\rightarrow 0}h^{-2s}\pi^2\epsilon_1\epsilon_2\int
d{\bf y}~e^{i({\bf x}-{\bf x}_0)\cdot J\cdot\tilde{{\bf y}}/\hbar}\cdot\nonumber \\
&&W({\bf x}+{\bf y}/2;E_n,\epsilon_1)W^*({\bf x}_0-{\bf y}/2;E_m,\epsilon_2)
\label{centid}
\end{eqnarray}
if the energy eigenvalues $E_n$ and $E_m$ are both nondegenerate,
 the case for a chaotic system.
In the limit of small  $\epsilon_1,\epsilon_2$ we obtain
\begin{eqnarray}
&&({\bf x}|n,m)(n,m|{\bf x}_0)\sim 
h^{-2s}\pi^2\epsilon_1\epsilon_2\int
d{\bf y}~e^{i({\bf x}-{\bf x}_0)\cdot J\cdot\tilde{{\bf y}}/\hbar}\cdot \nonumber \\
&&W({\bf x}+{\bf y}/2;E_n,\epsilon_1)W^*({\bf x}_0-{\bf y}/2;E_m,\epsilon_2)
\label{eq1}
\end{eqnarray}
which is amenable to semiclassical analysis. 

As a first approximation
we neglect the scar corrections and employ 
\begin{equation}
W({\bf x};E,\epsilon) \sim \delta_{\epsilon}(E-H({\bf x})) = h^{-1}
\int_{-\infty}^{\infty} dt ~e^{i(E-H({\bf x}))t/\hbar} e^{-\epsilon|t|/\hbar}
\label{eq:10}
\end{equation}
as we did in the stationary case. 
Substituting Eq. (\ref{eq:10}) into Eq. (\ref{eq1}), followed by a
simple change of variables ${\bf y}\rightarrow h {\bf y}$, yields
\begin{eqnarray}
&&({\bf x}|n,m)(n,m|{\bf x}_0)\sim
\pi^2\epsilon_1\epsilon_2\int
d{\bf y}~e^{2\pi i({\bf x}-{\bf x}_0)\cdot J\cdot\tilde{{\bf y}}}
\delta_{\epsilon_1}(E_n-H({\bf x}+h{\bf
y}/2))\delta_{\epsilon_2}(E_m-H({\bf x}_0-h{\bf y}/2)) \nonumber \\
&&=\frac{\pi^2\epsilon_1\epsilon_2}{h^2}\int
d{\bf y}dt_1dt_2e^{-\epsilon_1|t_1|/\hbar}e^{-\epsilon_2|t_2|/\hbar}e^{2\pi i({\bf x}-{\bf x}_0)\cdot J\cdot\tilde{{\bf y}}}\cdot
\nonumber \\
&&\exp\{i(E_n-H({\bf x}+h{\bf y}/2))t_1/\hbar\}
\exp\{i(E_m-H({\bf x}_0-h{\bf y}/2))t_2/\hbar\}.
\label{eq3}
\end{eqnarray}
Note, at this stage, the presence of essential singularities in each of the highly
oscillatory phase factors
$\exp\{i(E_n-H({\bf x}+h{\bf y}/2))t_1/\hbar\}$ and 
$\exp\{i(E_m-H({\bf x}_0-h{\bf y}/2))t_2/\hbar\}$.

We now let $h\rightarrow 0$ with
$\epsilon_1,~\epsilon_2 $ fixed.  Expanding
 the displaced Hamiltonian functions in Eq. (\ref{eq3})
in powers of $h$:
\begin{equation}
H({\bf x}+h{\bf y}/2)\sim
H({\bf x})+\frac{h}{2}\frac{\partial H({\bf x})}{\partial {\bf
x}}\cdot\tilde{{\bf y}}\sim H({\bf x})-\frac{h}{2}[\frac{\partial H({\bf x})}{\partial {\bf
x}}J]\cdot J\cdot\tilde{{\bf y}}
\label{h1}
\end{equation}
\begin{equation}
H({\bf x}_0-h{\bf y}/2)\sim
H({\bf x}_0)-\frac{h}{2}\frac{\partial H({\bf x}_0)}{\partial {\bf
x}_0}\cdot\tilde{{\bf y}}\sim H({\bf x}_0)+\frac{h}{2}[\frac{\partial H({\bf x}_0)}{\partial {\bf
x}_0}J]\cdot J\cdot \tilde{{\bf y}}
\label{h2}
\end{equation}
and substituting these expressions into Eq. (\ref{eq3}) gives,
\begin{eqnarray}
&&({\bf x}|n,m)(n,m|{\bf x}_0)\sim
\frac{\pi^2\epsilon_1\epsilon_2}{h^2}\int
d{\bf
y}dt_1dt_2e^{-\epsilon_1|t_1|/\hbar}e^{-\epsilon_2|t_2|/\hbar}
e^{i(E_n-H({\bf x}))t_1/\hbar}
e^{i(E_m-H({\bf x}_0))t_2/\hbar}\cdot\nonumber \\
&&\exp\{2\pi i\cdot [({\bf x}-\frac{\partial
H({\bf x})}{\partial {\bf x}}Jt_1/2)-({\bf x}_0-\frac{\partial
H({\bf x}_0)}{\partial {\bf x}_0}Jt_2/2)]\cdot J\cdot \tilde{{\bf y}}\}.
\label{eq4}
\end{eqnarray}
The factor $e^{-\epsilon_1|t_1|/\hbar}e^{-\epsilon_2|t_2|/\hbar}$
guarantees that the integrand is zero for all but short times since
$\epsilon_1|t_1|/\hbar\gg 2\pi|t_1|/T_{\rm min}$ and $\epsilon_2|t_2|/
\hbar\gg 2\pi|t_2|/T_{\rm min}$, so that we can 
use the short time approximation:
\begin{equation}
{\bf X}({\bf x},-t_1/2)\sim {\bf x}-\frac{\partial
H({\bf x})}{\partial {\bf x}}Jt_1/2,
\label{t1}
\end{equation}
\begin{equation}
{\bf X^{\prime}}({\bf x}_0,-t_2/2)\sim {\bf x}_0-\frac{\partial
H({\bf x}_0)}{\partial {\bf x}_0}Jt_2/2.
\label{t3}
\end{equation}
Using these results in Eq. (\ref{eq4}) gives
\begin{eqnarray}
&&({\bf x}|n,m)(n,m|{\bf x}_0)\sim
\frac{\pi^2\epsilon_1\epsilon_2}{h^2}\int
d{\bf
y}dt_1dt_2e^{-\epsilon_1|t_1|/\hbar}e^{-\epsilon_2|t_2|/\hbar}e^{i(E_n-H({\bf
x}))t_1/\hbar}\cdot 
\nonumber \\
&&e^{i(E_m-H({\bf x}_0))t_2/\hbar}
\exp\{2\pi i[{\bf X}({\bf x},-t_1/2)-{\bf
X^{\prime}}({\bf x}_0,-t_2/2)]\cdot J\cdot \tilde{{\bf y}}\} \nonumber\\
&&=\frac{\pi^2\epsilon_1\epsilon_2}{h^2}\int
dt_1dt_2e^{-\epsilon_1|t_1|/\hbar}e^{-\epsilon_2|t_2|/\hbar}\cdot 
\nonumber \\
&&e^{i(E_n-H({\bf
x}))t_1/\hbar}e^{i(E_m-H({\bf x}_0))t_2/\hbar}\delta({\bf X}({\bf x},-t_1/2)-{\bf X^{\prime}}({\bf x}_0,-t_2/2)).
\label{eq6}
\end{eqnarray}
Noting that Eq. (\ref{eq6}) is identically zero unless
${\bf x}$ and ${\bf x}_0$ are on the same
trajectory and using the fact that the Hamiltonian is time
independent allows us to replace $H({\bf x})$ by $H({\bf x}_0)$ in the 
exponential. Next
we perform a canonical transformation (i.e., time
translation) to put Eq. (\ref{eq6}) in the form
\begin{eqnarray}
&&({\bf x}|n,m)(n,m|{\bf x}_0)\sim
\frac{\pi^2\epsilon_1\epsilon_2}{h^2}\int
dt_1dt_2e^{-\epsilon_1|t_1|/\hbar}e^{-\epsilon_2|t_2|/\hbar}\cdot 
\nonumber \\
&&e^{i(E_n-H({\bf
x}_0))t_1/\hbar}e^{i(E_m-H({\bf x}_0))t_2/\hbar}
\delta({\bf x}_0-{\bf X}({\bf x},-(t_1-t_2)/2)) \nonumber\\
&&=\frac{\pi^22\epsilon_1\epsilon_2}{h^2}\int
 dt^{\prime}dt_0e^{-\epsilon_1|t_0|/\hbar}e^{-\epsilon_2|t_0-2t^{\prime}|/\hbar}\cdot \nonumber\\
&&e^{i(E_n-H({\bf 
x}_0))t_0/\hbar}e^{i(E_m-H({\bf x}_0))(t_0-2t^{\prime})/\hbar} 
\delta({\bf x}_0-{\bf X}({\bf x},-t^{\prime})).
\label{eq8}
\end{eqnarray}
where we have changed variables to $t^{\prime}=(t_1-t_2)/2$ and
$t_0=t_1$.
Note that $e^{-\epsilon_1|t_0|/\hbar}\sim 0$ unless $t_0\sim 0$ so
that we can replace $e^{-\epsilon_2|t_0-2t^{\prime}|/\hbar}$ by
$e^{-2\epsilon_2|t'|/\hbar}$ in Eq. (\ref{eq8}). Defining $\epsilon_0=\epsilon_1/2$ and
$\epsilon^{\prime}=2\epsilon_2$ we obtain
\begin{eqnarray}
&&({\bf x}|n,m)(n,m|{\bf x}_0)\sim
\frac{\pi^22\epsilon_0\epsilon^{\prime}}{h^2}\int
dt^{\prime}dt_0e^{-2\epsilon_0|t_0|/\hbar}e^{-\epsilon^{\prime}|t^{\prime}|/\hbar}\nonumber
\\
&& e^{i(E_n-H({\bf x}_0))t_0/\hbar}e^{i(E_m-H({\bf
x}_0))(t_0-2t^{\prime})/\hbar}
\delta({\bf x}_0-{\bf X}({\bf x},-t^{\prime})).
\end{eqnarray}
Using identity (\ref{ids}) in reverse then yields two equivalent forms:
\begin{equation}
({\bf x}|n,m)(n,m|{\bf x}_0)\sim
\frac{\pi^2\epsilon_0\epsilon^{\prime}}{h}\delta_{\epsilon_0}(E_{n,m}-H({\bf x}_0))\int
dt^{\prime}e^{-\epsilon^{\prime}|t^{\prime}|/\hbar} e^{-2i(E_m-H({\bf x}_0))t^{\prime}/\hbar}\delta({\bf x}_0-{\bf
X}({\bf x},-t^{\prime}))
\label{eq9}
\end{equation}
and
\begin{equation}
({\bf x}|n,m)(n,m|{\bf x}_0)\sim
\frac{\pi^2\epsilon_0\epsilon^{\prime}}{h}\delta_{\epsilon_0}(E_{n,m}-H({\bf x}_0))\int
dt^{\prime}e^{-\epsilon^{\prime}|t^{\prime}|/\hbar}
e^{i\lambda_{n,m}t^{\prime}}\delta({\bf x}_0-{\bf X}({\bf x},-t^{\prime})).
\label{eq10}
\end{equation}

Note that Eq. (\ref{eq10}) no longer exhibits
the essential singularities present in Eq. (\ref{eq3}). This is due to
the expansions
in Eqs. (\ref{h1}) and (\ref{h2}) through which the essential
singularities are eliminated. We found that the
same mechanism, i.e., elimination of essential singularities, was
responsible for correspondence in chaotic mappings of the torus
\cite{IVwbcat}.

We now take the limit ``$T\rightarrow\infty$'', that is we let
$\epsilon_0/\hbar, \epsilon^{\prime}/\hbar
\rightarrow 0$.
Note that $\epsilon_0$ and $\epsilon'$ essentially define a cutoff in
time beyond which the semiclassical approximations break down.
The commonly adopted cutoff time is the density of states
time $T_{{\rm ds}}\sim h/\langle \Delta E\rangle$. Since our $t'$ is symmetric
about zero, propagation to $T_{{\rm ds}}$ implies that 
$-T_{{\rm ds}}/2\leq
t'\leq T_{{\rm ds}}/2$. To achieve this we let $\epsilon'T_{{\rm
ds}}/2\hbar\rightarrow 1$ or, substituting $T_{{\rm ds}}\sim h/\langle
\Delta E\rangle$, $\pi \epsilon'\rightarrow \langle
\Delta E\rangle$. Thus the precise limits we must take to achieve the
``$T\rightarrow \infty$'' limit
$\epsilon_0/\hbar,\epsilon^{\prime}/\hbar\rightarrow
0$ are $\pi\epsilon_0,\pi\epsilon^{\prime}\rightarrow \langle
\Delta E\rangle$. The relation between the $\epsilon_0/\hbar,
\epsilon^{\prime}/\hbar\rightarrow
0$ limit and the $T_{{\rm ds}}\rightarrow \infty$ limit is explicit in a
formula proven by Kay\cite{IVkay}:
\begin{equation}
\lim_{\epsilon^{\prime}/\hbar\rightarrow
0}\epsilon^{\prime}/\hbar\int
dt^{\prime}e^{-\epsilon^{\prime}|t^{\prime}|/\hbar}\cdot=\lim_{T\rightarrow\infty}\frac{2}{T}\int_{-T/2}^{T/2}dt^{\prime}\cdot.
\label{ilim}
\end{equation}

We consider the correspondence limit of
Eq. (\ref{eq10}) for the case of $n\neq m$, as well as for $n=m$.
Consider first $n\neq m$. Performing the limits as
outlined above and making use of Eq. (\ref{ilim}) we obtain
\begin{eqnarray}
 ({\bf x}| n, m)(n,m|{\bf x}_0)&& =\lim_{\epsilon_0/\hbar\rightarrow 0} \frac{\pi \epsilon_0}{2}
\delta_{\epsilon_0} (E_{n,m}-H({\bf x}_0))
\lim_{\epsilon^{\prime}/\hbar\rightarrow 0}\frac{\epsilon^{\prime}}{\hbar}\int
dt^{\prime}e^{-\epsilon^{\prime}|t^{\prime}|/\hbar}e^{i\lambda_{n,m}t^{\prime}}\delta({\bf
x}_0-{\bf X}({\bf x},-t^{\prime}))\nonumber\\
&& =\lim_{\epsilon_0/\hbar\rightarrow 0} \frac{\pi \epsilon_0}{2}
\delta_{\epsilon_0} (E_{n,m}-H({\bf x}_0))
\lim_{T\rightarrow\infty}\frac{2}{T}\int_{-T/2}^{T/2}dt^{\prime}e^{i\lambda_{n,m}t^{\prime}}\delta({\bf x}_0-{\bf X}({\bf x},-t^{\prime}))\nonumber \\
&&=\lim_{\epsilon_0/\hbar\rightarrow 0} \frac{\pi \epsilon_0}{2}
\delta_{\epsilon_0} (E_{n,m}-H({\bf x}_0))
2\frac{d\lambda}{2\pi} \int_{-\infty}^{\infty}dt^{\prime}e^{i\lambda_{n,m}
t^{\prime}}\delta({\bf x}_0-{\bf X}({\bf x},-t^{\prime})),\nonumber \\
&&=dEd\lambda\delta(E_{n,m}-H(({\bf x}_0))\frac{1}{2\pi}
\int_{-\infty}^{\infty}dt'e^{i\lambda_{n,m}t} \delta({\bf x}_0-{\bf X}({\bf
x},-t^{\prime})
\end{eqnarray}
Here we have interpreted the limit of $1 / T$ to be $d\lambda / 2\pi$ and
$\pi \epsilon_0=\langle \Delta E\rangle\approx dE$. To
see that this is correct recall that we have taken the limit as
$\pi\epsilon '\rightarrow \langle \Delta E\rangle$ which
corresponds to letting $T\rightarrow T_{\rm ds}$. The inverse of the
 density of states
time can be roughly interpreted as the average nearest neighbor
Liouville frequency divided by $2\pi$. The average nearest neighbor
Liouville frequency $\langle\Delta E\rangle/\hbar$ can
evidently be interpreted as $d\lambda$. Given Eq.
 (\ref{eq26}) we have 
 \begin{equation}
 \lim_{h\rightarrow 0}~({\bf x} |n, m)(n,m|{\bf x}_0)=dE d\lambda~
 \Upsilon_{E_{n,m},\lambda_{n,m}}({\bf x};{\bf x}_0),
 \label{eq71}
 \end{equation}
 hence proving correspondence.

Note that in establishing Eq. (\ref{eq71}) we have also shown that individual quantum Liouville
eigenstates $({\bf x}|n,m)$, with $n\neq m$, do not have correspondence
limits when the classical system is chaotic. We may also infer the
reason: individual states $({\bf x}|n,m)$, $n\neq m$, possess
essential singularities which cancel in the product $({\bf
x}|n,m)(n,m|{\bf x}_0)$ to give a well defined correspondence limit.

Consider now the case where $n=m$ in Eq. (\ref{eq10}).  Using
Eq. (\ref{ilim}) and the fact that the classical dynamics is ergodic, we can
replace the time average by a phase average and rewrite the integral
in Eq. (\ref{eq10}) as
\begin{eqnarray}
&&\lim_{\epsilon^{\prime}/\hbar\rightarrow 0}\epsilon^{\prime}/\hbar\int
dt^{\prime}e^{-\epsilon^{\prime}|t^{\prime}|/\hbar}~\delta({\bf
x}_0-{\bf X}({\bf x},-t^{\prime})) 
=\lim_{T\rightarrow\infty}\frac{2}{T}\int_{-T/2}^{T/2}dt^{\prime}~\delta({\bf x}_0-{\bf X}({\bf x},-t^{\prime}))\nonumber \\
&&=2\int
d{\bf x}_0~\frac{\delta(E-H({\bf x}_0))}{\int d{\bf x}'\delta
(E-H({\bf x}'))}\delta({\bf x}_0-{\bf X}({\bf x},-t^{\prime}))\nonumber
\\
&&=2\frac{\delta(E-H({\bf x}))}{\int d{\bf x}'\delta(E-H({\bf x}'))}.
\end{eqnarray}
Substituting this expression into Eq. (\ref{eq10}) and again
interpreting $\pi\epsilon_0\sim \langle\Delta E\rangle \sim dE$ we
obtain [Eq. (\ref{statc})],  the desired correspondence:
\begin{equation}
\lim_{h\rightarrow 0}
~({\bf x}|n,n)(n,n|{\bf x}_0)=dE~ \Upsilon_{E_n}({\bf x};{\bf x}_0).
\end{equation}
Since 
\begin{equation}
\Upsilon_{E}({\bf x};{\bf x}_0)=\rho_E^*({\bf
x}_0)\rho_E({\bf x})=({\bf x}|E)(E|{\bf x}_0),
\label{eq73}
\end{equation}
Eq. (\ref{eq73}) implies that a product of stationary quantum
Liouville eigenfunctions goes to a product of stationary classical
Liouville eigenfunctions. However, this is not the case for the
nonstationary projectors.  That is,
\begin{equation}
\Upsilon_{E,\lambda}({\bf x};{\bf
x}_0)=\sum_{\ell}\rho_{E,\lambda}^{\ell *}({\bf x}_0)\rho_{E,\lambda}^{\ell} ({\bf
x}) = \sum_{\ell}~ ({\bf x}|E,\lambda,\ell)(E,\lambda,\ell|{\bf x}_0)
\end{equation}
is not a simple product of Liouville eigenfunctions, but rather
a sum of products. Thus Eq. (\ref{eq26}) implies that the
correspondence limit of a product of
nonstationary quantum Liouville eigenfunctions is a sum of
products of nonstationary classical Liouville eigenfunctions, due to
the degeneracy of the classical states. 

\subsection{Correspondence for the Liouville Spectrum}
\label{4.4.2}

In addition to the limit relations for the spectral projection
operators discussed in Sec. \ref{4.4.1}, correspondence requires that the
quantum spectrum reduce to its classical analog in the 
$h\rightarrow 0$ limit. Since the classical Liouville spectrum is continuous
we examine spectral
densities, rather than individual Liouville eigenvalues. 
The quantum Liouville operator $\delta(\lambda-\hat{L})$ can be
expanded on the Liouville eigenbasis as:
\begin{equation}
\delta(\lambda-\hat{L})=\sum_{n,m}|n,m)(n,m|\delta(\lambda-\hat{L})|n,m)(n,m|.
\end{equation}
The trace $D(\lambda)$ of $\delta(\lambda-\hat{L})$,
 is the quantum Liouville spectral density, i.e.,
\begin{eqnarray}
D(\lambda) = {\rm Tr}[\delta(\lambda-\hat{L})] &=& \sum_{n,m}~
(n,m|\delta(\lambda-\hat{L})|n,m)  \nonumber \\
&=& \sum_{n,m}~(n,m|\delta(\lambda-\lambda_{n,m})|n,m)
\nonumber \\
&=& \sum_{n,m}\delta(\lambda-\lambda_{n,m}).
\label{ldens}
\end{eqnarray}

With a view toward investigating the classical limit we note that we
can rewrite the first equality of Eq. (\ref{ldens}), by inserting the 
identity $\delta(\lambda-\hat{L})=(2\pi)^{-1}\int dt ~e^{i(\lambda-\hat{L})t}$, as
\begin{equation}
D(\lambda)=\sum_{n,m}\frac{1}{2\pi}\int_{-\infty}^{\infty}dt
~e^{i\lambda t}~(n,m|e^{-i\hat{L}t}|n,m).
\label{ldens2}
\end{equation}
But, inserting the closure relation (\ref{xclosure}), and noting that $\hat{L}|{\bf x})=L({\bf x})|{\bf x})$, gives
\begin{equation}
(n,m|e^{-i\hat{L}t}|n,m)=\int d{\bf
x}~(n,m|{\bf x})~e^{-iL({\bf x})t}({\bf x}|n,m).
\end{equation}
It follows that Eq. (\ref{ldens2}) can be
rewritten in the form
\begin{eqnarray}
D(\lambda)&=&\sum_{n,m}
\frac{1}{2\pi}\int_{-\infty}^{\infty}dt ~e^{i\lambda
t}\int d{\bf x}~(n,m|{\bf x})~e^{-iL({\bf x})t}({\bf x}|n,m)\nonumber \\
&=&
\frac{1}{2\pi}\int_{-\infty}^{\infty}dt ~e^{i\lambda
t}\int d{\bf x}d{\bf x}_0\left[\sum_{n,m}~(n,m|{\bf
x})({\bf x}_0|n,m)\right]e^{-iL({\bf x})t}\delta({\bf x}-{\bf x}_0)\nonumber \\
&=&
\frac{1}{2\pi}\int_{-\infty}^{\infty}dt ~e^{i\lambda
t}\int d{\bf x}d{\bf x}_0\delta({\bf x}-{\bf x}_0)e^{-iL({\bf x})t}\delta({\bf
x}-{\bf x}_0),
\end{eqnarray}
where we have used $\sum_{m,n}~(n,m|{\bf
x})({\bf x}_0|n,m)=\delta({\bf x}-{\bf x}_0)$. Formally expanding
$L({\bf x})$ in powers of Planck's constant and taking the $h\rightarrow 0$
limit, gives
$L({\bf x})\rightarrow L_c({\bf x})$, where
$L_c({\bf x})=iH({\bf x})\sigma$ is the classical Liouville operator. It
follows that as $h\rightarrow 0$
\begin{eqnarray}
D(\lambda)&&\rightarrow\frac{1}{2\pi}\int_{-\infty}^{\infty}dt ~e^{i\lambda
t}\int d{\bf x}d{\bf x}_0~\delta({\bf x}-{\bf x}_0)e^{-iL_c({\bf x})t}\delta({\bf
x}-{\bf x}_0)\nonumber \\
&&\rightarrow\frac{1}{2\pi}\int_{-\infty}^{\infty}dt ~e^{i\lambda
t}\int d{\bf x}d{\bf x}_0~\delta({\bf x}-{\bf x}_0)\delta({\bf
X}({\bf x},-t)-{\bf x}_0)\nonumber \\
&&\rightarrow\frac{1}{2\pi}\int_{-\infty}^{\infty}dt ~e^{i\lambda
t}\int d{\bf x}~\delta({\bf
X}({\bf x},-t)-{\bf x}).
\end{eqnarray}
But 
\begin{eqnarray}
\frac{1}{2\pi}\int_{-\infty}^{\infty}dt ~e^{i\lambda
t}\int d{\bf x}~\delta({\bf
X}({\bf x},-t)-{\bf x})&=&\int_0^{\infty} dE\int_{-\infty}^{\infty}
d\lambda_0~\delta(\lambda-\lambda_0)\int d{\bf
x}~\Upsilon_{E,\lambda}({\bf x};{\bf x})\nonumber \\
&\equiv & D_c(\lambda)
\end{eqnarray} 
where $D_c(\lambda)$ is the classical Liouville spectral density\cite{IVeckh}. 

Thus, in a formal sense we have correspondence, i.e.,
$D(\lambda)\rightarrow D_c(\lambda)$, in the $h\rightarrow 0$ limit. 
However, the proof is unsatisfactory because the classical spectrum is
 highly degenerate and this limit is not well defined, i.e., 
 $D_c(\lambda)=\infty$.
This arises from the fact that the
classical Liouville spectrum is infinitely degenerate due to its
stability with respect to variations with energy\cite{IVeckh,IVhannay}.
In addition, this formal proof provides little insight into the way
that the spectra approach one another.

This problem can be bypassed by considering the Liouville spectral
density for energies in a classically small, but quantum mechanically
large, energy interval $E_0-\epsilon/2\leq E \leq E_0+\epsilon/2 $.
That is, we define
\begin{eqnarray}
D_{\epsilon}(E_0;\lambda)&=&\sum_{n,m\atop E_0-\epsilon/2\leq
E_{n,m}\leq E_0+\epsilon/2}\delta
(\lambda-\lambda_{n,m})\nonumber \\
&=&\hbar\int_{E_0-\epsilon/2}^{E_0+\epsilon/2}dE~
d(E+\hbar\lambda/2)d(E-\hbar\lambda/2),
\label{elspec}
\end{eqnarray}
where $d(E)=\sum_n \delta (E-E_n)$.
Then, as shown below, expression (\ref{elspec}) has a well
defined classical limit, i.e.,
\begin{equation}
\lim_{\epsilon\rightarrow 0}\lim_{h\rightarrow
0}D_{\epsilon}(E_0;\lambda)=D_c(E_0;\lambda)
\label{slim}
\end{equation}
with
\begin{equation}
D_c(E_0;\lambda)=\delta(\lambda)+ \frac{1}{2\pi}\int_{-\infty}^{\infty}dt ~e^{i\lambda
t}\int_{H({\bf x})=E_0}d{\bf x}~\delta ({\bf
X}({\bf x},-t)-{\bf x}).
\label{cspect2}
\end{equation}
Here $D_c(E_0;\lambda)$ is the classical Liouville density of
states\cite{IVhannay,IVnota} on the energy surface
$H({\bf x})=E_0$.  Note that the factor $\delta ({\bf X}({\bf x},-t)-{\bf
x})$ in Eq. (\ref{cspect2}) is nonzero only for points ${\bf x}$ which
lie on periodic orbits of period $t$. Thus the integral $\int_{H({\bf
x})=E_0}d{\bf x}~\delta ({\bf
X}({\bf x},-t)-{\bf x})$ can be written as a sum over periodic
orbits\cite{IVhannay}, giving \cite{IVeckh,IVhannay}
\begin{equation}
D_c(E_0;\lambda)=\delta(\lambda)+\frac{1}{\pi}\sum_{j}\frac{T_j(E_0)\cos
(\lambda T_j(E_0))}{k_j|{\rm
det}(M_j(E_0)-I)|},
\label{csum}
\end{equation}
where $T_j$ is the period of periodic orbit $j$, $k_j$ is its
winding number, $M_j$ is its $2s-2\times 2s-2$ dimensional 
stability matrix, and
the sum is over positive traversals of the periodic orbits.

To show Eq. (\ref{slim}) 
we employ Gutzwiller's formula 
$d(E)=\bar{d}(E)+d_{{\rm osc}}(E)$ for the density of states, in Eq.
(\ref{elspec}).  Here  $\bar{d}(E)=\langle\Delta
E\rangle^{-1}$ is the average density of states, and $d_{{\rm osc}}(E)$ is an
oscillatory correction given by the formula
\begin{equation}
d_{{\rm osc}}(E)\sim \sum_{j}d^{j}_{{\rm osc}}(E)
\label{posum}
\end{equation}
where
\begin{equation}
d_{{\rm osc}}^{j}(E)\sim \frac{T_j(E)}{k_j\pi\hbar}\frac{\cos
(S_j(E)/\hbar+\gamma_j)}{\sqrt{|{\rm det}(M_j(E)-I)|}}.
\label{djk}
\end{equation}
Here $S_j(E)$ is the action of the periodic orbit $j$ and
$\gamma_j=\sigma_j \pi/2$, where $\sigma_j$ is the Maslov index
of the orbit\cite{IVdelos,IVcrl,IVsc}.
The sum in Eq. (\ref{posum}) is over positive
traversals of the periodic orbits.

The Gutzwiller formula for the density of states is not generally
convergent, but
can be made so by broadening over energy\cite{IVberryd3}.  That is, we
replace $d(E)$ by the
energy broadened density $d_{\mu}(E)=\sum_n\Omega_{\mu}(E-E_n)$ where
$\Omega_{\mu}(x)=1/\mu$ for $-\mu/2\leq x\leq \mu/2$ and is zero otherwise.
Note that $\lim_{\mu\rightarrow 0}\Omega_{\mu}(x)=\delta(x)$.
The energy broadening modifies the standard Gutzwiller expansion by
damping out contributions from very long periodic orbits. 

We rewrite Eq. (\ref{elspec}) in the form
\begin{equation}
D_{\epsilon}(E_0;\lambda)=\lim_{\mu\rightarrow \langle\Delta E\rangle}\hbar\int_{E_0-\epsilon/2}^{E_0+\epsilon/2}dE 
~d_{\mu}(E+\hbar\lambda/2)d_{\mu}(E-\hbar\lambda/2)
\label{spec2}
\end{equation}
in order to employ the energy broadened (and hence
convergent\cite{IVconv})
form of Gutzwiller's formula. The
correspondence limit is now $h\rightarrow 0$ followed by
$\epsilon,\mu\rightarrow \langle \Delta E\rangle$\cite{footpi}.

We now separate the Liouville density $D_{\epsilon}(E_0;\lambda)$
into its diagonal and off-diagonal parts. Since the system is chaotic
we assume that it exhibits level repulsion, i.e., that the probability of two
neighboring energy levels exhibiting an accidental degeneracy is zero.
It therefore follows that the only contribution to the Liouville spectrum at
$\lambda=0$ is from the diagonal ($n=m$) terms,
\begin{equation}
\lim_{\lambda\rightarrow 0}D_{\epsilon}(E_0;\lambda)\sim \lim_{\mu\rightarrow \langle\Delta E\rangle}\hbar\int_{E_0-\epsilon/2}^{E_0+\epsilon/2}dE 
~\sum_{n}\Omega_{\mu}(E+\hbar\lambda/2-E_n)\Omega_{\mu}(E-\hbar\lambda/2-E_n),
\end{equation}
so that we can  rewrite Eq. (\ref{spec2}) in the limit $h\rightarrow 0$ 
(i.e., $h\lambda\rightarrow 0$ for all $\lambda$) in the form
\begin{eqnarray}
D_{\epsilon}(E_0;\lambda)&\sim &\lim_{\mu\rightarrow \langle\Delta E\rangle}\hbar\int_{E_0-\epsilon/2}^{E_0+\epsilon/2}dE 
~\sum_{n}\Omega_{\mu}(E+\hbar\lambda/2-E_n)\Omega_{\mu}(E-\hbar\lambda/2-E_n)\nonumber \\
&+&\lim_{\mu\rightarrow \langle\Delta E\rangle}\hbar\int_{E_0-\epsilon/2}^{E_0+\epsilon/2}dE 
~[d_{\mu}(E+\hbar\lambda/2)d_{\mu}(E-\hbar\lambda/2)-d_{\mu}^2(E)].
\label{spec3}
\end{eqnarray}
Note that as $\mu\rightarrow \langle\Delta E\rangle$
\begin{equation}
\sum_{n}
\Omega_{\mu}(E+\hbar\lambda/2-E_n)\Omega_{\mu}(E-\hbar\lambda/2-E_n)
\rightarrow \delta (\hbar\lambda)~d(E)
\end{equation}
and so the first term in Eq. (\ref{spec3}) becomes
\begin{eqnarray}
&&\lim_{\mu\rightarrow \langle\Delta E\rangle}\hbar\int_{E_0-\epsilon/2}^{E_0+\epsilon/2}dE 
~\sum_{n}
\Omega_{\mu}(E+\hbar\lambda/2-E_n)\Omega_{\mu}(E-\hbar\lambda/2-E_n)\nonumber
\\
&&\sim
\hbar\epsilon\delta(\hbar\lambda)\frac{1}{\epsilon}\int_{E_0-\epsilon/2}^{E_0+\epsilon/2}dE
~d(E).
\end{eqnarray}
If we choose $\epsilon \sim h/T_{{\rm min}}$, where $T_{{\rm min}}$ is
the period of the shortest periodic orbit of energy $E_0$, then
\begin{equation}
\bar{d}(E_0)\sim \frac{1}{\epsilon}\int_{E_0-\epsilon/2}^{E_0+\epsilon/2}dE 
~d(E)
\end{equation}
and so we may write Eq. (\ref{spec3}) in the form 
\begin{equation}
D_{\epsilon}(E_0;\lambda)=
\hbar\epsilon\delta(\hbar\lambda)\bar{d}(E_0)+\lim_{\mu\rightarrow
\langle\Delta E\rangle}\hbar\int_{E_0-\epsilon/2}^{E_0+\epsilon/2}dE 
~[d_{\mu}(E+\hbar\lambda/2)d_{\mu}(E-\hbar\lambda/2)-d_{\mu}^2(E)].
\label{spec4}
\end{equation}
Now we assume that $d_{\mu}(E)\sim \bar{d}(E)+d_{{\rm osc},\mu}(E)$, with
$\frac{1}{\epsilon}\int_{E_0-\epsilon/2}^{E_0+\epsilon/2}dE
~d_{{\rm osc},\mu}(E)\sim 0$, and
substituting this expression into Eq. (\ref{spec4}) we obtain
\begin{eqnarray}
D_{\epsilon}(E_0;\lambda)&\sim&\hbar\epsilon\delta(\hbar\lambda)\bar{d}(E_0)+
\hbar\epsilon \bar{d}^{~2}(E_0)-\hbar\lim_{\mu\rightarrow
\langle\Delta
E\rangle}\int_{E_0-\epsilon/2}^{E_0+\epsilon/2}dE~d_{\mu}^2(E)\nonumber
\\
&+&\lim_{\mu\rightarrow \langle\Delta E\rangle}\hbar\int_{E_0-\epsilon/2}^{E_0+\epsilon/2}dE 
~d_{{\rm osc},\mu}(E+\hbar\lambda/2)d_{{\rm osc},\mu}(E-\hbar\lambda/2).
\label{spec5}
\end{eqnarray}
The form of the energy broadened density is now introduced. In
particular\cite{broad},
\begin{eqnarray}
d_{\mu}(E)&\sim &\bar{d}(E)+\sum_j\frac{T_j(E)~{\rm sinc}(\mu
T_j(E)/2\hbar)}{k_j\pi\hbar\sqrt{|{\rm det}(M_j(E)-I)|}}\cos
(S_j(E)/\hbar+\gamma_j)\nonumber \\
&\sim &\bar{d}(E)+\sum_jA_{\mu,j}(E)\cos
(S_j(E)/\hbar+\gamma_j),
\label{mugutz}
\end{eqnarray}
where 
\begin{equation}
A_{\mu,j}(E)\equiv \frac{T_j(E)~{\rm sinc}(\mu
T_j(E)/2\hbar)}{k_j\pi\hbar\sqrt{|{\rm det}(M_j(E)-I)|}},
\end{equation}
and ${\rm sinc}(x)=\sin(x)/x$ is the damping function.

With $d_{{\rm osc},\mu}(E)\sim \sum_jA_{\mu,j}(E)\cos(S_j(E)/\hbar+\gamma_j)$
the last  term of Eq. (\ref{spec5}) can be written as
\begin{eqnarray}
&&\int_{E_0-\epsilon/2}^{E_0+\epsilon/2}
dE~d_{{\rm osc},\mu}(E+\hbar\lambda/2)d_{{\rm osc},\mu}(E-\hbar\lambda/2)\nonumber\\
&&\sim
\sum_{j,j'}\int_{E_0-\epsilon/2}^{E_0+\epsilon/2}dE~
A_{\mu,j}(E+\hbar\lambda/2)A_{\mu,j'}(E-\hbar\lambda/2)\cdot\nonumber \\
&&\cos (S_j(E+\hbar\lambda/2)/\hbar+\gamma_j)\cos
(S_{j'}(E-\hbar\lambda/2)/\hbar+\gamma_{j'})
\nonumber \\
&\sim &
\sum_{j,j'}\int_{E_0-\epsilon/2}^{E_0+\epsilon/2}dE~
A_{\mu,j}(E+\hbar\lambda/2)A_{\mu,j'}(E-\hbar\lambda/2)\cdot\nonumber \\
&&\{ \cos
\left[(S_j(E+\hbar\lambda/2)-S_{j'}(E-\hbar\lambda/2))/\hbar+\gamma_j-\gamma_{j'}\right]\nonumber\\
&+&\cos
\left[(S_j(E+\hbar\lambda/2)+S_{j'}(E-\hbar\lambda/2))/\hbar+\gamma_j+\gamma_{j'}\right]\}
\label{eqyuk}
\end{eqnarray}
Now let $h\rightarrow 0$, followed by
$\epsilon,\mu\rightarrow \langle\Delta E\rangle$.
Observing that $\Omega_{\mu}(E-E_n)\Omega_{\mu}(E-E_m)\sim
\Omega_{\mu}(E-E_n)\delta_{n,m}/\mu$ if $\mu$ is sufficiently small
(a consequence of level repulsion),
if follows that $d_{\mu}(E)^2\sim d_{\mu}(E)/\mu$. Taking
$\mu\rightarrow\langle\Delta E\rangle$ we see that
\begin{equation}
\int_{E_0-\epsilon/2}^{E_0+\epsilon/2}dE~d_{\mu}^2(E)\sim \int_{E_0-\epsilon/2}^{E_0+\epsilon/2}dE~d(E)/\langle \Delta
E\rangle \sim \epsilon \bar{d}^{~2}(E_0).
\end{equation}
Substituting this result back into Eq. (\ref{spec5}) we obtain the
result that
\begin{equation}
D_{\epsilon}(E_0;\lambda)\sim \hbar\epsilon\delta(\hbar\lambda)\bar{d}(E_0)+
\lim_{\mu\rightarrow \langle\Delta E\rangle}\hbar\int_{E_0-\epsilon/2}^{E_0+\epsilon/2}dE 
~d_{{\rm osc},\mu}(E+\hbar\lambda/2)d_{{\rm osc},\mu}(E-\hbar\lambda/2).
\label{bigyuk}
\end{equation}

It remains to evaluate Eq. (\ref{eqyuk}), a complicated procedure
outlined in Appendix B.  Then
taking the limit as $\epsilon\rightarrow \langle\Delta
E\rangle$, and substituting the contributions from Eqs. (\ref{cont1}),
(\ref{cont2}),
(\ref{cont3}), and (\ref{cont4}) into Eq. (\ref{eqyuk}) and
substituting Eq. (\ref{eqyuk}) back into Eq. (\ref{bigyuk}) 
gives the result:
\begin{eqnarray}
D_{\epsilon}(E_0;\lambda)&\sim & \delta (\lambda)+ \frac{1}{\pi}\sum_{j}\frac{T_j(E_0)\cos
(\frac{S_j(E_0+\lambda\hbar/2)-S_j(E_0-\hbar\lambda/2)}{\hbar})}{k_j|{\rm
det}(M_j(E_0)-I)|}\nonumber \\
&+& O(h^{s-1}e^{iz/\hbar}).
\label{pspeclim}
\end{eqnarray}
For $s\geq 2$, we thus see that
\begin{equation}
\lim_{\epsilon\rightarrow 0}\lim_{h\rightarrow 0}D_{\epsilon}(E_0;\lambda)=\delta(\lambda)+\frac{1}{\pi}\sum_{j}\frac{T_j(E_0)\cos
(\lambda T_j(E_0))}{k_j|{\rm
det}(M_j(E_0)-I)|}=D_{c}(E_0;\lambda).
\label{speclim}
\end{equation}
That is, the quantum Liouville spectrum properly approaches the classical
Liouville spectrum as $h\rightarrow 0$. Note that Eq. (\ref{speclim})
emerges from Eq. (\ref{pspeclim}) via elimination of essential
singularities. 

\section{Scar Corrections}
\label{4.5}

In the last section we began with  Eq. (\ref{eq1}) and utilized Berry's 
formula [Eq. (\ref{bf})] for 
$W({\bf x};E,\epsilon)$, neglected scar corrections and arrived at a proof
of Eqs. (\ref{lim1}) and (\ref{lim2}), i.e., the correspondence rules
discussed above. 
We now consider the corrections to these limits which arise due
to the scars from the periodic orbits.

Consider first the following formula for $W^j_{{\rm scar}}({\bf x};E,\epsilon)$,
the scar contribution to $W({\bf x};E,\epsilon)$ from periodic orbit $j$:
\begin{eqnarray}
&&W_{{\rm
scar}}^j({\bf x};E,\epsilon)\sim \frac{2^s}{\sqrt{|{\rm det}(M_j+I)|}}e^{-\epsilon
T_j/\hbar}\cdot\nonumber \\
&&\cos\{[S_j-\mbox{\boldmath
$\xi$}[J(M_j-I)/(M_j+I)]\tilde{\mbox{\boldmath $\xi$}}]/\hbar+\gamma_j\}\cdot
\nonumber \\
&&h^{-1}\int_{-\infty}^{\infty}dt ~e^{-\epsilon
|t|/\hbar}e^{i\{(E-H)t-\frac{1}{24}\dot{{\bf x}}\wedge\ddot{{\bf
x}}t^3\}}.
\label{scar1}
\end{eqnarray}
Here the variables $\mbox{\boldmath $\xi$}$ are the $2(s-1)$
coordinates
of the surface of section transverse to the periodic orbit, and $\dot{{\bf x}}\wedge\ddot{{\bf
x}}= |\nabla V({\bf q})|^2/m +({\bf p}\cdot\nabla)^2V({\bf q})/m^2$.
The derivation of Eq. (\ref{scar1}) 
is given by Berry\cite{IVberry3} although
he does not write it out explicitly.

Let $h \rightarrow 0$ with $\epsilon \gg h/T_{{\rm min}}$.
[For convenience we drop the $j$ subscripts on
$M_j,S_j,\gamma_j$ and $T_j$.] 
Note that the $t^3$ term in the time integral of Eq. (\ref{scar1}) can
be neglected because
only short times count for small $h$. Thus
\begin{eqnarray}
&&W_{{\rm
scar}}^j({\bf x};E,\epsilon)\sim \frac{2^s}{\sqrt{|{\rm det}(M+I)|}}e^{-\epsilon
T/\hbar}\cdot\nonumber \\
&&\cos\{[S(E)-\mbox{\boldmath $\xi$}[J(M-I)/(M+I)]\tilde{\mbox{\boldmath $\xi$}}]/\hbar+\gamma\}\delta_{\epsilon}(E-H({\bf x})) \nonumber \\
&&= \frac{2^s}{\sqrt{|{\rm det}(M+I)|}}e^{-\epsilon
T/\hbar}
\delta_{\epsilon}(E-H({\bf x}))[e^{i\{[S(E)-\mbox{\boldmath $\xi$}[J(M-I)/(M+I)]\tilde{\mbox{\boldmath $\xi$}}]/\hbar+\gamma\}}\nonumber \\
&&+e^{-i\{[S(E)-\mbox{\boldmath $\xi$}[J(M-I)/(M+I)]\tilde{\mbox{\boldmath $\xi$}}]/\hbar+\gamma\}}]/2
\label{cos}
\end{eqnarray}
Consider now factors like $e^{\pm i\mbox{\boldmath
$\xi$}[J(M-I)/(M+I)]\tilde{\mbox{\boldmath $\xi$}}/\hbar}$ in the $h\rightarrow 0$
limit. Note that the integral
\begin{equation}
\int d{\bf z}~e^{\pm i{\bf z}\Omega\tilde{{\bf z}}/\alpha}f({\bf z})
\end{equation}
[where ${\bf z}=(z_1,\dots,z_N)$ and $\Omega$ is an $N\times N$ dimensional matrix
independent of ${\bf z}$] in the limit $\alpha\rightarrow 0$ can be
evaluated by stationary phase to give
\begin{equation}
\int d{\bf z}~e^{\pm i{\bf z}\Omega\tilde{{\bf z}}/\alpha}f({\bf z})\sim
\frac{[\alpha\pi]^{N/2}}{\sqrt{|\det (\Omega)|}}f(0)e^{\pm i\pi{\rm sgn}\Omega/4}.
\end{equation}
This suggests the existence of a distributional identity
\begin{equation}
e^{\mp i\pi{\rm sgn}\Omega/4}\frac{\sqrt{|\det (\Omega)|}}{[\alpha\pi]^{N/2}}e^{\pm i{\bf z}\Omega\tilde{{\bf
z}}/\alpha}\rightarrow \delta({\bf z}).
\end{equation}
This formula, when applied to the exponent in Eq. (\ref{cos}), gives
\begin{equation}
e^{\pm i\mbox{\boldmath $\xi$}[J(M-I)/(M+I)]\tilde{\mbox{\boldmath $\xi$}}/\hbar}\rightarrow e^{\pm i\sigma }(h/2)^{s-1}\sqrt{|\frac{\det (M+I)}{\det
(M-I)}|}\delta(\mbox{\boldmath $\xi$})
\label{dident}
\end{equation}
[here $\sigma=\pi{\rm sgn}[J(M-I)/(M+I)]/4$].  Substituting into Eq. (\ref{cos}) 
 gives
\begin{equation}
W_{{\rm
scar}}^j({\bf x};E,\epsilon)\sim \frac{2h^{s-1}}{\sqrt{|{\rm det}(M-I)|}}e^{-\epsilon
T/\hbar}\cos\{S(E)/\hbar+\gamma\}\delta_{\epsilon}(E-H({\bf
x}))\delta (\mbox{\boldmath $\xi$}({\bf x})),
\label{elyuko}
\end{equation}
which is the classical limit obtained by
Berry\cite{IVberry3}.  

To obtain the scar corrections to the Liouville spectral projectors we
insert Eq. (\ref{bf}) into Eq. (\ref{eq1}) and use Eq.
(\ref{elyuko})
for the scar term.  Their are two types of corrections of the forms:
``$\delta_{\epsilon}(E-H)\times$~scar''  and
``scar~$\times$~scar''.  The latter are
of much higher order in $h$ and are neglected. The correction terms
 to Eq.
(\ref{eq3}) due to periodic
orbits of period $T$ and energy $E_n$ are then 
\begin{eqnarray}
&&{\cal S}_{n,m}({\bf x};{\bf x}_0)=\frac{2\pi^2\epsilon_1\epsilon_2h^{s-2}}{\sqrt{|{\rm
det}(M-I)|}}e^{-\epsilon_1T/\hbar}\cos\{S(E_n)/\hbar
+\gamma\}\cdot\nonumber \\
&&\int
d{\bf
y}dt_1dt_2e^{-\epsilon_1|t_1|/\hbar}e^{-\epsilon_2|t_2|/\hbar}e^{2\pi
i({\bf x}-{\bf x}_0)\cdot J\cdot\tilde{{\bf y}}}\delta(\mbox{\boldmath $\xi$}({\bf x}+h{\bf y}/2))\cdot
\nonumber \\
&&\exp\{i(E_n-H({\bf x}+h{\bf y}/2))t_1/\hbar\}\exp\{i(E_m-H({\bf x}_0-h{\bf y}/2))t_2/\hbar\}.
\label{mucker}
\end{eqnarray}
We neglect $h$ corrections to $\mbox{\boldmath $\xi$}$, i.e., we assume that
$\mbox{\boldmath $\xi$}({\bf x}+h{\bf y}/2)\sim
\mbox{\boldmath $\xi$}({\bf x})$. This can be justified as follows: (a)
expanding $\mbox{\boldmath $\xi$}({\bf x}+h{\bf y}/2)$
to first order in $h$ and using
the fact that $J(M-I)/(M+I)$ is symmetric\cite{IVberry3}, and Eq. 
(\ref{dident}), allows us to
show that
\begin{equation}
\delta(\mbox{\boldmath $\xi$}({\bf x}+h{\bf
y}/2))\sim\delta(\mbox{\boldmath $\xi$}({\bf
x}))\cos\{2\pi[{\bf y}\cdot \frac{\partial \mbox{\boldmath $\xi$}}{\partial {\bf x}}]\cdot[J(M-I)/(M+I)]\cdot \tilde{\mbox{\boldmath  
$\xi$}}\},
\label{muck}
\end{equation}
and
(b) noting that the right hand side of Eq. (\ref{muck}) is zero
unless $\mbox{\boldmath $\xi$}({\bf x})\sim 0$, and that the argument of the
cosine factor is proportional to $\mbox{\boldmath $\xi$}({\bf x})$
allows us to replace the cosine factor by unity.

Using the short time expansions of Eqs. (\ref{t1}) and (\ref{t3}) and 
doing the integrals over
${\bf u}$ and ${\bf v}$ gives:
\begin{eqnarray}
&&{\cal S}_{n,m}({\bf x};{\bf x}_0)=\frac{2\pi^2\epsilon_1\epsilon_2h^{s-2}}{\sqrt{|{\rm
det}(M-I)|}}e^{-\epsilon_1T/\hbar}\cos\{S(E_n)/\hbar
+\gamma\}\delta(\mbox{\boldmath $\xi$}({\bf x}))\cdot\nonumber \\
&&\int
dt_1dt_2e^{-\epsilon_1|t_1|/\hbar}e^{-\epsilon_2|t_2|/\hbar}
\exp\{i(E_n-H({\bf x}))t_1/\hbar\}\cdot\nonumber
\\
&&\exp\{i(E_m-H({\bf x}_0))t_2/\hbar\}\delta({\bf X}({\bf x},-t_1/2)-{\bf X^{\prime}}({\bf x}_0,-t_2/2)).
\end{eqnarray}
Changing variables to $t'=(t_1-t_2)/2$ and $t_0=t_1$ and defining
$\epsilon_0=\epsilon_1/2$ and $\epsilon^{\prime}=2\epsilon_2$ 
as in Sec. \ref{4.4.1} we obtain:
\begin{eqnarray}
&&{\cal S}_{n,m}({\bf x};{\bf x}_0)\sim\frac{\pi\epsilon_0h^{s-1}}{2\sqrt{|{\rm
det}(M-I)|}}e^{-\epsilon_0T/\hbar}\cos\{S(E_n)/\hbar
+\gamma\}\delta_{\epsilon_0}(E_{n,m}-H({\bf x}))\cdot\nonumber \\
&&\delta(\mbox{\boldmath $\xi$}({\bf x}))\lim_{\epsilon '/\hbar\rightarrow 0}\epsilon '/\hbar\int
dt^{\prime}e^{-\epsilon '|t^{\prime}|/\hbar}
\exp\{i\lambda_{n,m}t^{\prime}\}\delta({\bf x}_0-{\bf X}({\bf x},-t^{\prime})).
\label{eq114}
\end{eqnarray}

For the stationary case $n=m$ we interchange limits via Eq.
(\ref{ilim}) to obtain:
\begin{eqnarray}
{\cal S}_{n,n}({\bf x};{\bf x}_0)&=&\frac{\langle\Delta
E\rangle^2}{h\sqrt{|{\rm
det}(M-I)|}}e^{-\epsilon_0T/\hbar}\cos\{S(E_n)/\hbar
+\gamma\}\cdot\nonumber \\
&&\delta(E_{n}-H({\bf x}))\delta(E_{n}-H({\bf
x}_0))\delta(\mbox{\boldmath $\xi$}({\bf x})). 
\end{eqnarray}
The other cross term, due to periodic orbits of energy $E_m$, gives a
similar contribution ${\cal S}_{n,n}({\bf x}_0;{\bf x})$, and so the
overall correction to Eq. (\ref{eq3}) with $n=m$, due to the 
periodic orbits, is (see also Berry \cite{IVberry3})
\begin{eqnarray}
&&{\cal S}_{n,n}({\bf x};{\bf x}_0)+{\cal S}_{n,n}({\bf x}_0;{\bf x})\sim \sum_{j}\frac{\langle\Delta E\rangle^2}{h\sqrt{|{\rm
det}(M_j-I)|}}e^{-\epsilon_0T_j/\hbar}\cos\{S_j(E_n)/\hbar
+\gamma_j\}\cdot\nonumber \\
&&\delta(E_{n}-H({\bf x}_0))\delta(E_{n}-H({\bf x}))\left[\delta(\mbox{\boldmath $\xi$}_j({\bf x}))+\delta(\mbox{\boldmath $\xi$}_j({\bf x}_0))\right].
\label{eq116}
\end{eqnarray}
This result may be rewritten in terms of the distributions

\begin{equation}
\Upsilon_{E,0}^j({\bf x};{\bf x}_0)=\frac{k_j}{T_j}\delta(E-H({\bf x}_0))\delta(E-H({\bf x}))\delta(\mbox{\boldmath $\xi$}_j({\bf x})).
\label{eq117}
\end{equation}
which we defined in Paper 1 \cite{paper1}. and which are stationary
spectral projectors with uniform density on the periodic orbits. Using
Eq. (\ref{eq117}), Eq. (\ref{eq116}) becomes  
\begin{eqnarray}
&&{\cal S}_{n,n}({\bf x};{\bf x}_0)+{\cal S}_{n,n}({\bf x}_0;{\bf x})
=\sum_{j}\frac{T_j\langle\Delta E\rangle^2}{k_jh\sqrt{|{\rm
det}(M_j-I)|}}e^{-\epsilon_0T_j/\hbar}\cos\{S_j(E_n)/\hbar
+\gamma_j\}\cdot\nonumber \\
&&\left[\Upsilon_{E_n,0}^j({\bf x};{\bf x}_0)+\Upsilon_{E_n,0}^j({\bf
x}_0;{\bf x})\right] .
\label{po1}
\end{eqnarray}
This result shows that the scar corrections to the limit as $h
\rightarrow 0$ of the spectral projectors corresponding to stationary
states are comprised of weighted sums over the stationary classical
projectors which have uniform density on the classical periodic orbits.
For the nonstationary case $n\neq m$, again interchanging limits via
Eq. (\ref{ilim}) and considering only points on the periodic orbit, it
follows that the integral in Eq. (\ref{eq114}) can be written as:
\begin{eqnarray}
&&\lim_{\epsilon '/\hbar\rightarrow 0}\epsilon '/\hbar\int
dt^{\prime}e^{-\epsilon '|t^{\prime}|/\hbar}
e^{i\lambda_{n,m}t^{\prime}}
\delta({\bf x}_0-{\bf X}({\bf x},-t^{\prime}))\nonumber \\
&&=\lim_{k\rightarrow\infty}\frac{1}{2k+1}\frac{\sin\{(2k+1)\lambda_{n,m}\tau/2\}}{\sin(\lambda_{n,m}\tau/2)}\cdot\nonumber
\\
&&\frac{2}{\tau}\int_{-\tau/2}^{\tau/2}
dt^{\prime}e^{i\lambda_{n,m}t^{\prime}}\delta({\bf x}_0-{\bf X}({\bf x},-t^{\prime})).
\label{eq119}
\end{eqnarray}
[Here $\tau=T_j/k_j$.]
Since it can be readily shown that
\[
\lim_{k\rightarrow\infty}\frac{1}{2k+1}\frac{\sin\{(2k+1)\lambda_{n,m}\tau/2\}}{\sin(\lambda_{n,m}\tau/2)}
=\left\{ \begin{array}{ll} 
1 & \mbox{if
$ \lambda_{n,m}=2\pi l/\tau$ for $l\in {\bf Z}$}\\
0 & \mbox{otherwise}
\end{array}
\right. \]
it follows that there are scar
corrections only for the nonstationary distributions whose frequency
matches an integer multiple of the frequency of one of the periodic
orbits. When this
condition is met the scar correction to the nonstationary spectral
projector $(n,m|{\bf x})({\bf x}_0|n,m)$ is obtained from Eqs.
(\ref{eq114}) and (\ref{eq119}) as
\begin{eqnarray}
&&{\cal S}_{n,m}({\bf x};{\bf x}_0)\sim \frac{\langle\Delta E\rangle h^{s-1}}{\sqrt{|{\rm
det}(M-I)|}}e^{-\epsilon_0T/\hbar}\cos\{S(E_n)/\hbar
+\gamma\}\delta_{\epsilon_0}(E_{n,m}-H({\bf x}))\cdot\nonumber \\
&&\delta(\mbox{\boldmath $\xi$}({\bf
x}))\frac{1}{\tau}\int_{-\tau/2}^{\tau/2}
dt^{\prime}e^{i\lambda_{n,m}t^{\prime}}\delta({\bf x}_0-{\bf
X}({\bf x},-t^{\prime})).
\label{nssc1}
\end{eqnarray}
Similar corrections ${\cal S}_{m,n}({\bf x}_0;{\bf x})$ arise from
periodic orbits of energy $E_m$, i.e., ${\cal S}_{n,m}({\bf x};{\bf
x}_0)$ is given by Eq.(\ref{nssc1}) with $E_n$ replaced by $E_m$.  

Consider now these nonstationary corrections in more detail.
The scar contribution [Eq. (\ref{nssc1})] involves
the following product of factors 
\begin{equation}
\delta_{\epsilon_0}(E_{n,m}-H({\bf x}))\delta(\mbox{\boldmath
$\xi$}({\bf x})).
\label{prd}
\end{equation}
The variables $\mbox{\boldmath $\xi$}({\bf x})$ are effectively zero
on a local family of periodic orbits with energies close to $E_n$ and
periods close to $T$. The distribution
$\delta(\mbox{\boldmath $\xi$}({\bf x}))$ is thus zero except
on this local family. If the energy $E_{n,m}$ lies outside of the
neighborhood in energy of the local family then the product of delta
functions [Eq. (\ref{prd})] will be everywhere zero. As a
consequence the product is generally zero for $n\neq m$. The same
considerations
hold for the scar term $S_{n,m}({\bf x};{\bf x}_0)$ with periodic
orbits of energy $E_m$. Thus, scar corrections to the nonstationary
Liouville eigenfunctions are typically negligible in the semiclassical
limit. Only periodic orbits of period $T_j/k_j=2\pi l/\lambda_{n,m}$,
$l\in{\bf Z}$, contribute and of these only the ones with energy $E_n$
or $E_m$ close to $E_{n,m}$ make a nonzero contribution.

Thus, we see that the stationary and nonstationary contributions of periodic
orbits [Eqs. (\ref{po1}) and (\ref{nssc1})] at most make corrections of order
$h^{2s-1}e^{-\epsilon T/\hbar}$ which vanish in the correspondence limit 
($h\rightarrow 0$ followed by $\pi\epsilon\rightarrow \langle\Delta E\rangle$).
Furthermore, in the classical limit these corrections are only supported 
on the measure zero set of periodic orbits.

\section{Correspondence: Applications to Matrix Elements}
\label{4.6}

The results obtained above allow us to systematize and extend previous results
on the classical limiting
forms of matrix elements of quantum observables.
With our normalization of the quantum Liouville eigenfunctions,
matrix elements satisfy the following relationship
\begin{equation}
\langle n|\hat{A}|m\rangle=h^{-s/2}
\int d{\bf x}~\rho_{n,m}^w({\bf x})~A^w({\bf x}). 
\end{equation}
As a consequence we may relate matrix elements of observables to the
spectral projection operators via the expression:
\begin{equation}
|\langle n|\hat{A}|m\rangle|^2=\int d{\bf x}d{\bf
x}_0~A^{w*}({\bf
x})({\bf x}|n,m)(n,m|{\bf x}_0)A^{w}({\bf x}_0).
\label{junk}
\end{equation}
We focus on the correspondence limit of matrix elements for
observables $\hat{A}$ with a well
defined classical analog, i.e., for $A^w({\bf x})\rightarrow A({\bf x})$
where $A({\bf x})$ is the classical analog.

Consider first that Eqs. (\ref{junk}), (\ref{eq26}), and (\ref{nm}) 
imply that $(n\neq m)$
\begin{equation}
\lim_{h\rightarrow 0}
|\langle n|\hat{A}|m\rangle|^2=\frac{\langle\Delta
E\rangle}{h}\int_{-\infty}^{\infty}dt'e^{i\lambda_{n,m}t'}~\langle
A(0)A(t')\rangle_{E_{n,m}}
\end{equation}
where $\langle
A(0)A(t')\rangle_{E_{n,m}}$ denotes the microcanonical average of
$A({\bf x},0)A({\bf x},t')$ at an energy $E_{n,m}$.
Noting that
\begin{eqnarray}
&&\int_{-\infty}^{\infty}dt'e^{i\lambda_{n,m}t'}~\langle
A(0)A(t')\rangle_{E_{n,m}}=2\pi\langle
A\rangle_{E_{n,m}}^2\delta(\lambda_{n,m})+\nonumber\\
&&\int_{-\infty}^{\infty}dt'e^{i\lambda_{n,m}t'}~\langle
[A(0)-\langle A\rangle_{E_{n,m}} ][A(t')-\langle
A\rangle_{E_{n,m}}]\rangle_{E_{n,m}}\\
&&=\int_{-\infty}^{\infty}dt'e^{i\lambda_{n,m}t'}~\langle
[A(0)-\langle A\rangle_{E_{n,m}} ][A(t')-\langle
A\rangle_{E_{n,m}}]\rangle_{E_{n,m}},
\end{eqnarray}
if $\lambda_{n,m}\neq 0$, it follows that
\begin{equation}
\lim_{h\rightarrow 0}|\langle n|\hat{A}|m\rangle|^2=\frac{\langle\Delta
E\rangle}{h}\int_{-\infty}^{\infty}dt'e^{i\lambda_{n,m}t'}~\langle
[A(0)-\langle A\rangle_{E_{n,m}} ][A(t')-\langle A\rangle_{E_{n,m}}]\rangle_{E_{n,m}}
\label{ndiag}
\end{equation}
for the nonstationary case. This result is essentially in agreement
with the predictions of Feingold and Peres\cite{IVperes}. Their result
differs from Eq. (\ref{ndiag}) insofar as they are
missing a factor of $\hbar^{-1}$,
and they left the energy $E_{n,m}$ unspecified. 

The arguments
leading to Eq. (\ref{ndiag}) can be readily generalized to the case
of mixed operators with the result that
\begin{equation}
\lim_{h\rightarrow 0}\langle m|\hat{A}|n\rangle\langle n|\hat{B}|m\rangle
= \frac{\langle\Delta
E\rangle}{h}\int_{-\infty}^{\infty}dt'e^{i\lambda_{n,m}t'}~\langle
[A(0)-\langle A\rangle_{E_{n,m}} ][B(t')-\langle
B\rangle_{E_{n,m}}]\rangle_{E_{n,m}}.
\label{abe}
\end{equation}
This important result establishes a connection between the Liouville
spectrum of classical time correlation functions and quantum matrix
elements for chaotic systems.

The case of $n=m$ also results from Eq. (\ref{junk}) and 
Eq. (\ref{statc}) to give the well known results:
\begin{eqnarray}
\lim_{h\rightarrow 0}|\langle n|\hat{A}|n\rangle|^2 &=& \langle A\rangle_{E_n}^2
\nonumber \\
\lim_{h\rightarrow 0}\langle n|\hat{A}|n\rangle\langle n|\hat{B}|n\rangle &
=& \langle
A\rangle_{E_n}\langle B\rangle_{E_n}.
\label{mix}
\end{eqnarray}

Scar corrections to the off-diagonal matrix elements are negligible.
Scar corrections for the stationary case $|\langle
n|\hat{A}|n\rangle|^2$ are of the form
\begin{equation}
\sum_{j}\frac{2T_j\langle\Delta E\rangle}{k_jh\sqrt{|{\rm
det}(M_j-I)|}}e^{-\epsilon_0T_j/\hbar}\cos\{S_j(E_n)/\hbar
+\gamma_j\}~\langle A\rangle_{E_n}~\langle A\rangle_{E_n,0}^j,
\label{correct}
\end{equation}
where 
\begin{equation}
\langle
A\rangle_{E,0}^j\equiv\frac{1}{\tau_j}\int_{-\tau_j/2}^{\tau_j/2}dt'~A({\bf
X}({\bf x}^j,t'))
\end{equation}
is the average of $A$ over the periodic orbit $j$ and ${\bf x}^j$ is a
point on the periodic orbit,
while corrections to Eq. (\ref{mix}) are of the form
\begin{equation}
\sum_{j}\frac{T_j\langle\Delta E\rangle}{k_jh\sqrt{|{\rm
det}(M_j-I)|}}e^{-\epsilon_0T_j/\hbar}\cos\{S_j(E_n)/\hbar
+\gamma_j\}[~\langle A\rangle_{E_n}~\langle B\rangle_{E_n,0}^j+\langle
B\rangle_{E_n}~\langle A\rangle_{E_n,0}^j].
\end{equation}
Comparing Eq. (\ref{ndiag}) to Eq. (\ref{correct}) it is readily
apparent that squared off-diagonal matrix elements for a chaotic system
are of the same order of magnitude as the fluctuations in the squared diagonal
matrix elements, i.e., $O(h^{s-1})$. This confirms an early conjecture
by Pechukas\cite{IVpech}.

Finally, we compare the chaotic case to the well known \cite{IVjaffe2},
\cite{IVwardlaw}  classical
limit for matrix elements of individual matrix elements for integrable
systems, i.e.,
\begin{equation}
\lim_{h\rightarrow 0}\langle {\bf n}|\hat{A}|{\bf m}\rangle =\frac{1}{(2\pi)^s}\int d\mbox{\boldmath $\theta$} ~A({\bf I}_{{\bf n},{\bf m}},\mbox{\boldmath $\theta$})~e^{i({\bf
n}-{\bf m})\cdot\mbox{\boldmath $\theta$}}.
\label{bj1}
\end{equation}
The most significant difference is that Eq. (\ref{bj1}), which holds for both the stationary
and nonstationary integrable cases, gives a classical limit for individual matrix
elements whereas the chaotic results allow classical limits only for
products of matrix elements [Eqs. (\ref{ndiag}), (\ref{abe}) and
(\ref{mix})].
This is a direct consequence of the fact that $\rho_{n,m}$ has a
classical limit for regular systems but not for chaotic systems.

\section{Summary}
\label{4.7}

Correspondence for chaotic quantum systems has been considered from
the viewpoint of distribution dynamics in the Wigner-Weyl
representation of quantum mechanics. The connections between quantum
and classical dynamics have been clarified through the formulation of
the correspondence problem in terms of
Liouville spectral projection operators. Our
demonstration of correspondence for these objects and for the Liouville
spectrum shows that quantum dynamics is capable
of reproducing chaotic classical dynamics in the $h\rightarrow 0$
limit. The mechanism of correspondence here, as in our studies of
chaotic mappings, appears to be the elimination of
essential singularities. Corrections arising from periodic orbits were
also considered with the result that
stationary quantum spectral projection operators have
corrections in the form of weighted sums of stationary classical projectors on
periodic orbits. Scar corrections for the nonstationary Liouville
spectral projection operators were found to be negligible. 
Applications of our correspondence results to matrix elements revealed
connections between
the matrix elements of quantum observables
and the spectrum of classical time correlation functions. The
corrections to the diagonal matrix elements due to periodic orbits
were shown to be of the same order of magnitude in $h$ as the
square of the off-diagonal matrix elements. 

\vspace{1 in}
{\bf Acknowledgements:} We thank the Natural Sciences and Engineering
Research Council of Canada for support of this work. PB acknowledges
stimulating discussions with Professor J. Ford, Georgia Institute of
Technology.

\pagebreak

\section*{Appendix A}
\label{appendixA}

We prove identity (\ref{subid2p}) by showing that the righthand
side can be reduced to the left. To begin we substitute the
definitions of $({\bf x}|n,n)$ into the righthand side to obtain
\begin{eqnarray}
&&h^{-2s}\int d{\bf u}d{\bf v}d{\bf q}'d{\bf q}'' ~e^{-i({\bf p}-{\bf
p}_0)\cdot{\bf v}/\hbar}e^{i({\bf q}-{\bf
q}_0)\cdot{\bf u}/\hbar}e^{i({\bf p}+{\bf
u}/2)\cdot{\bf q}'/\hbar}e^{i({\bf p}_0-{\bf
u}/2)\cdot{\bf q}''/\hbar}\cdot\nonumber \\
&&\langle {\bf q}+{\bf v}/2-{\bf q}'/2|n\rangle\langle n|{\bf q}+{\bf
v}/2+{\bf q}'/2\rangle\cdot\nonumber \\
&&\langle {\bf q}_0-{\bf v}/2-{\bf
q}''/2|m\rangle\langle m|{\bf q}_0-{\bf
v}/2+{\bf q}''/2\rangle.
\end{eqnarray}
Performing the integral over ${\bf u}$ gives
\begin{eqnarray}
&&h^{-s}\int d{\bf v}d{\bf q}'d{\bf q}''~\delta({\bf q}-{\bf q}_0+{\bf
q}'/2-{\bf q}''/2)e^{-i({\bf p}-{\bf
p}_0)\cdot{\bf v}/\hbar}e^{i{\bf p}\cdot{\bf q}'/\hbar}e^{i{\bf p}_0\cdot{\bf q}''/\hbar}\cdot\nonumber \\
&&\langle {\bf q}+{\bf v}/2-{\bf q}'/2|n\rangle\langle n|{\bf q}+{\bf
v}/2+{\bf q}'/2\rangle\cdot\nonumber \\
&&\langle {\bf q}_0-{\bf v}/2-{\bf
q}''/2|m\rangle\langle m|{\bf q}_0-{\bf
v}/2+{\bf q}''/2\rangle.
\end{eqnarray}
Now doing the integral over ${\bf q}''$ gives
\begin{eqnarray}
&&h^{-s}2^s\int d{\bf v}d{\bf q}'~e^{i{\bf p}\cdot(-{\bf v}+{\bf
q}')/\hbar}e^{-i{\bf p}_0\cdot(2{\bf q}_0-2{\bf q}-{\bf v}-{\bf
q}')/\hbar}\cdot\nonumber\\
&&\langle {\bf q}-(-{\bf v}+{\bf q}')/2|n\rangle\langle m|{\bf q}+(-{\bf
v}+{\bf q}')/2\rangle\cdot\nonumber\\
&&\langle {\bf q}_0+(2{\bf q}_0-2{\bf
q}-{\bf v}-{\bf q}')/2|m\rangle\langle n|{\bf q}_0 -(2{\bf q}_0-2{\bf q}-{\bf v}-{\bf q}')/2\rangle.
\end{eqnarray}
Changing variables to ${\bf v}'=-{\bf v}+{\bf q}'$, and ${\bf
v}''=2{\bf q}_0-2{\bf q}-{\bf v}-{\bf q}'$ gives a Jacobian
determinant of absolute value $1/2^s$. Inserting the definitions of
$({\bf x}|n,m)$ we immediately obtain the righthand side of Eq.
(\ref{subid2p}).

\section*{Appendix B}
\label{appendixB}

Here we systematically evaluate each of the contributions to Eq.
(\ref{eqyuk}), in accord with the correspondence limit $h\rightarrow 0$,
$\epsilon\gg h/T_{{\rm min}}$, followed by
$\epsilon\rightarrow\langle\Delta E\rangle$. We assume that $T_j(E)$ and
$M_j(E)$ vary slowly with energy, i.e., $T_{j}(E\pm\hbar\lambda/2)\sim
T_{j}(E)$ and $M_{j}(E\pm\hbar\lambda/2)\sim M_{j}(E)$ for small $h$, and note, as a
consequence, that the amplitudes $A_{\mu,j}$ are also slowly varying,
i.e.,
\begin{equation}
A_{\mu,j}(E+\hbar\lambda/2)\sim A_{\mu,j}(E),
\end{equation}
and
\begin{equation}
A_{\mu,j'}(E+\hbar\lambda/2)\sim A_{\mu,j'}(E).
\end{equation}
We thus define the slowly varying amplitudes
\begin{equation}
B_{j,j'}(E)\equiv A_{\mu,j}(E)A_{\mu,j'}(E).
\end{equation}
The integrals to be evaluated [Eq. (\ref{eqyuk})] therefore take the form
\begin{equation}
\int_{E_0-\epsilon/2}^{E_0+\epsilon/2}dE~B_{j,j'}(E)\cos\left[\frac{S_j(E+\hbar\lambda/2)\pm
S_{j'}(E-\hbar\lambda/2)}{\hbar}+\gamma_j\pm\gamma_{j'}\right],
\label{evaluate}
\end{equation}
and we must consider three separate cases.

{\em Case 1}: We begin by
considering the integrals for which the actions add. Here, because we
are summing over positive traversals,
\begin{equation}
\frac{d}{dE}[S_j(E+\hbar\lambda/2)+S_{j'}(E-\hbar\lambda/2)]=T_j(E+\hbar\lambda/2)+T_{j'}(E-\hbar\lambda/2)>0
\end{equation}
and so there are no stationary phase points in the interval
$E_0-\epsilon/2\leq E\leq E_0+\epsilon/2$. It can be shown\cite{IVvain} that
integrals of the form $\int_a^bdz~f(z)e^{i\phi(z)/\hbar}$, with $f(z)$
slowly varying and $\phi'(z)\neq 0$ for $z\in [a,b]$, can be
approximated by
\begin{equation}
\int_a^bdz~f(z)e^{i\phi(z)/\hbar}\sim \frac{\hbar}{i}\left[
\frac{f(b)}{\phi'(b)}e^{i\phi(b)/\hbar}-\frac{f(a)}{\phi'(a)}e^{i\phi(a)/\hbar}\right]
\label{nspp}
\end{equation}
in the limit that $h\rightarrow 0$. 
Application of Eq. (\ref{nspp}) to the integral
\begin{equation}
\int_{E_0-\epsilon/2}^{E_0+\epsilon/2}dE~B_{j,j'}(E)\cos\left[\frac{S_j(E+\hbar\lambda/2)+
S_{j'}(E-\hbar\lambda/2)}{\hbar}+\gamma_j+\gamma_{j'}\right]
\end{equation}
gives
\begin{eqnarray}
&&\int_{E_0-\epsilon/2}^{E_0+\epsilon/2}
dE~B_{j,j'}(E)\cos\left[\frac{S_j(E+\hbar\lambda/2)+
S_{j'}(E-\hbar\lambda/2)}{\hbar}+
\gamma_j+\gamma_{j'}\right]\sim\nonumber\\ 
&&\hbar \{
\frac{B_{j,j'}(E_0+\epsilon/2)}{T_j(E_0+\epsilon/2)+T_{j'}(E_0+\epsilon/2)}
\cdot \nonumber
\\
&&\sin\left[\frac{S_j(E_0+\epsilon/2+\hbar\lambda/2)+
S_{j'}(E_0+\epsilon/2-\hbar\lambda/2)}{\hbar}+\gamma_j+
\gamma_{j'}\right] \nonumber \\
&&-\frac{B_{j,j'}(E_0-\epsilon/2)}{T_j(E_0-\epsilon/2)+T_{j'}(E_0-\epsilon/2)}
\cdot \nonumber
\\
&&\sin\left[\frac{S_j(E_0-\epsilon/2+\hbar\lambda/2)+
S_{j'}(E_0-\epsilon/2-\hbar\lambda/2)}{\hbar}+\gamma_j+\gamma_{j'}
\right] \}.
\label{yuk}
\end{eqnarray}
Taking the limit as $\epsilon\rightarrow \langle\Delta E\rangle$, 
$B_{j,j'}(E_0\pm\epsilon/2)\sim
B_{j,j'}(E_0)$ and $T_j(E_0\pm \epsilon/2)\sim T_j(E_0)$, so that we
can rewrite Eq. (\ref{yuk}) in the form
\begin{eqnarray}
&&\int_{E_0-\epsilon/2}^{E_0+\epsilon/2}
dE~B_{j,j'}(E)\cos\left[\frac{S_j(E+\hbar\lambda/2)+
S_{j'}(E-\hbar\lambda/2)}{\hbar}+
\gamma_j+\gamma_{j'}\right]\sim\nonumber\\ 
&&\frac{2\hbar B_{j,j'}(E_0)}{T_j(E_0)+T_{j'}(E_0)}
\cos\left[\frac{S_j(E_0+\hbar\lambda/2)+
S_{j'}(E_0-\hbar\lambda/2)}{\hbar}+\gamma_j+\gamma_{j'} \right]
\cdot\nonumber \\
&&\sin
[(T_j(E_0)+T_{j'}(E_0))\epsilon/\hbar]\nonumber \\
&&\sim 2\epsilon B_{j,j'}(E_0)
\cos\left[\frac{S_j(E_0+\hbar\lambda/2)+
S_{j'}(E_0-\hbar\lambda/2)}{\hbar}+\gamma_j+\gamma_{j'} \right]
\cdot\nonumber \\
&&{\rm sinc}
[(T_j(E_0)+T_{j'}(E_0))\epsilon/\hbar].
\label{restr1}
\end{eqnarray}
Further simplifications show that these contributions take the form
\begin{eqnarray}
&&\int_{E_0-\epsilon/2}^{E_0+\epsilon/2}
dE~B_{j,j'}(E)\cos\left[\frac{S_j(E+\hbar\lambda/2)+
S_{j'}(E-\hbar\lambda/2)}{\hbar}+
\gamma_j+\gamma_{j'}\right]\sim\nonumber\\ 
&&2\langle \Delta E\rangle B_{j,j'}(E_0)\cos
\left[\frac{S_j(E_0+\hbar\lambda/2)+
S_{j'}(E_0-\hbar\lambda/2)}{\hbar}+\gamma_j+\gamma_{j'} \right] .
\label{cont1}
\end{eqnarray}

Now consider the terms in Eq. (\ref{evaluate}) involving a difference of actions. 

{\em Case 2}: Of the terms with
$j\neq j'$, some will have
stationary phase points in the interval $E_0-\epsilon/2\leq E\leq
E_0+\epsilon/2$, and others will not. For those terms which do not have
stationary phase points
\begin{eqnarray}
&&\int_{E_0-\epsilon/2}^{E_0+\epsilon/2}
dE~B_{j,j'}(E)\cos\left[\frac{S_j(E+\lambda\hbar/2)-
S_{j'}(E-\hbar\lambda/2)}{\hbar}+
\gamma_j-\gamma_{j'}\right]\sim\nonumber\\ 
&&\frac{2\hbar B_{j,j'}(E_0)}{T_j(E_0)-T_{j'}(E_0)}
\cos\left[\frac{S_j(E_0+\hbar\lambda/2)-
S_{j'}(E_0-\hbar\lambda/2)}{\hbar}+\gamma_j-\gamma_{j'} \right]
\cdot\nonumber \\
&&\sin
[(T_j(E_0)-T_{j'}(E_0))\epsilon/\hbar]\nonumber \\
&&\sim 2\epsilon B_{j,j'}(E_0)
\cos\left[\frac{S_j(E_0+\hbar\lambda/2)-
S_{j'}(E_0-\hbar\lambda/2)}{\hbar}+\gamma_j-\gamma_{j'} \right]
\cdot\nonumber \\
&&{\rm sinc}
[(T_j(E_0)-T_{j'}(E_0))\epsilon/\hbar],
\label{restr2}
\end{eqnarray}
as one can easily show by applying Eq. (\ref{nspp}).
In the limit as $\epsilon\rightarrow \langle\Delta E\rangle$ we find that
\begin{eqnarray}
&&\int_{E_0-\epsilon/2}^{E_0+\epsilon/2}
dE~B_{j,j'}(E)\cos\left[\frac{S_j(E+\hbar\lambda/2)-
S_{j'}(E-\hbar\lambda/2)}{\hbar}+
\gamma_j-\gamma_{j'}\right]\sim\nonumber\\ 
&&2\langle \Delta E\rangle B_{j,j'}(E_0)
\sin \left[\frac{S_j(E_0+\hbar\lambda/2)-
S_{j'}(E_0-\hbar\lambda/2)}{\hbar}+\gamma_j-\gamma_{j'} \right].
\label{cont2}
\end{eqnarray}

Consider those terms which do have stationary phase points $E'\in
[E_0-\epsilon/2,E_0+\epsilon/2]$. We define
$\phi(E)=S_j(E+\hbar\lambda/2)-S_{j'}(E-\hbar\lambda/2)$. Expanding $\phi(E)$ about the
stationary phase point $E'$ gives $\phi(E)\sim
\phi(E')+\frac{1}{2}\phi''(E')(E-E')^2$. Now consider that
\begin{equation}
\int_{E_0-\epsilon/2}^{E_0+\epsilon/2}dE~B_{j,j'}(E)e^{i\phi(E)/\hbar}
\sim
B_{j,j'}(E')e^{i\phi(E')/\hbar}\int_{E_0-\epsilon/2}^{E_0+\epsilon/2}dE
~e^{i\phi''(E')(E-E')^2/2\hbar}.
\label{12many}
\end{equation}
This approximation resembles the stationary phase approximation\cite{IVvain} except
that we have retained the original limits on the remaining integral instead of
replacing them by $\pm\infty$. 
Evaluation of the remaining integral in Eq. (\ref{12many}) then yields \cite{IVabra}
\begin{eqnarray}
&&\int_{E_0-\epsilon/2}^{E_0+\epsilon/2}dE~B_{j,j'}(E)e^{i\phi(E)/\hbar}
\sim
\frac{1}{2}\sqrt{\frac{ih}{\phi''(E')}}B_{j,j'}(E')e^{i\phi(E')/\hbar}\cdot\nonumber
\\
&&\left[ {\rm
erf}~[\sqrt{\frac{-i\phi''(E')}{2\hbar}}(E_0+\epsilon/2-E')]-{\rm
erf}~[\sqrt{\frac{-i\phi''(E')}{2\hbar}}(E_0-\epsilon/2-E')] \right].~~~~~~~~
\end{eqnarray}
In the limit as $\epsilon\rightarrow \langle \Delta E\rangle$ we obtain
\begin{equation}
\int_{E_0-\epsilon/2}^{E_0+\epsilon/2}dE~B_{j,j'}(E)e^{i\phi(E)/\hbar}
\sim\langle \Delta E\rangle
B_{j,j'}(E')e^{i\phi(E')/\hbar}e^{i\phi ''(E')(E_0-E')^2/2\hbar},
\end{equation}
or
\begin{eqnarray}
&&\int_{E_0-\epsilon/2}^{E_0+\epsilon/2}
dE~B_{j,j'}(E)\cos\left[\frac{S_j(E+\hbar\lambda/2)-
S_{j'}(E-\hbar\lambda/2)}{\hbar}+\gamma_j-\gamma_{j'}\right]\nonumber
\\
&&\sim
\langle \Delta E\rangle
B_{j,j'}(E')\cos (\Phi),
\label{cont3}
\end{eqnarray}
where $\Phi=\phi(E')/\hbar+\phi ''(E')(E_0-E')^2/2\hbar +\gamma_j-\gamma_{j'}$.

{\em Case 3}: Finally, we consider the terms in Eq. (\ref{evaluate}) where $j=j'$. First note
that there are
no true stationary phase points, i.e.,
$T_j(E+\hbar\lambda/2)-T_j(E-\hbar\lambda/2)\neq 0$ for
$\lambda\neq 0$. Second, note that the action $S_j(E)$ changes by
$S_j(E_0+\epsilon/2)-S_j(E_0-\epsilon/2)\equiv \Delta S_j(E_0)\sim
T_j(E_0)\epsilon$ over the interval $E_0-\epsilon/2\leq E\leq
E_0+\epsilon/2$. Since $\epsilon\sim h/T_{{\rm min}}$ it follows that
$\Delta S_j(E_0)/\hbar\sim 2\pi T_j(E_0)/T_{{\rm min}}$. In general
$T_j(E_0)>>T_{{\rm min}}$ and so $S/\hbar$ changes dramatically over
the interval $E_0-\epsilon/2\leq E\leq E_0+\epsilon/2$. Similarly,
$\Delta [S_j(E_0+\hbar\lambda/2)-S_j(E_0-\hbar\lambda/2)]/\hbar\sim 2\pi
[T_j(E_0+\hbar\lambda/2)-T_j(E_0-\hbar\lambda/2)]/T_{{\rm min}}$, and
since even
$|T_j(E_0+\hbar\lambda/2)-T_j(E_0-\hbar\lambda/2)|>> T_{{\rm min}}$
in general, it
follows that the phase of the cosine factor will oscillate many times
over the interval $E_0-\epsilon/2\leq E\leq E_0+\epsilon/2$. By
contrast, over the interval $E_0-hk_j/2T_j(E_0)\leq E\leq E_0+hk_j/2T_j(E_0)$,
$\Delta
[S_j(E_0+\hbar\lambda/2)-S_j(E_0-\hbar\lambda/2)]/\hbar \sim 2\pi k_j
[T_j(E_0+\hbar\lambda/2)-T_j(E_0-\hbar\lambda/2)]/T_j(E_0)$. In
general
$|T_j(E_0+\hbar\lambda/2)-T_j(E_0-\hbar\lambda/2)|/T_j(E_0)<<1$ for
$h$ small and so the phase is effectively
stationary over the small interval $E_0-hk_j/2T_j(E_0)\leq E\leq E_0+hk_j/2T_j(E_0)$ about $E_0$, and we will therefore break
the integral into three parts:
\begin{eqnarray}
&&\int_{E_0-\epsilon/2}^{E_0+\epsilon/2}dE~B_{j,j}(E) \cos
(\frac{S_j(E+\lambda\hbar/2)-S_j(E-\hbar\lambda/2)}{\hbar})= \nonumber \\
&&[\int_{E_0-hk_j/2T_j(E_0)}^{E_0+hk_j/2T_j(E_0)}dE~
+\int_{E_0-\epsilon/2}^{E_0-hk_j/2T_j(E_0)}dE \nonumber \\
&&+\int_{E_0+hk_j/2T_j(E_0)}^{E_0+\epsilon/2}dE ]~
B_{j,j}(E)\cos
(\frac{S_j(E+\hbar\lambda/2)-S_j(E-\hbar\lambda/2)}{\hbar}).
\end{eqnarray}

The first term can be approximated via
\begin{eqnarray}
&&\int_{E_0-hk_j/2T_j(E_0)}^{E_0+hk_j/2T_j(E_0)}dE~B_{j,j}(E) \cos
(\frac{S_j(E+\hbar\lambda/2)-S_j(E-\hbar\lambda/2)}{\hbar})\sim \nonumber\\
&&\frac{k_jhB_{j,j}(E_0)}{T_j(E_0)} \cos
(\frac{S_j(E_0+\hbar\lambda/2)-S_j(E_0-\hbar\lambda/2)}{\hbar}),
\end{eqnarray}
while the other two terms can be evaluated\cite{IVvain} via Eq. (\ref{nspp}) to give 
\begin{eqnarray}
&&\int_{E_0-\epsilon/2}^{E_0-hk_j/2T_j(E_0)}dE~B_{j,j}(E) \cos
(\frac{S_j(E+\hbar\lambda/2)-S_j(E-\hbar\lambda/2)}{\hbar})\sim\nonumber \\
&&\frac{2\hbar
B_{j,j}(E_0)}{T_j(E_0+\hbar\lambda/2)-T_j(E_0-\hbar\lambda/2)} \cos
(\frac{S_j(E_0+\hbar\lambda/2)-S_j(E_0-\hbar\lambda/2)}{\hbar})\cdot\nonumber
\\
&&\sin \left[
(\epsilon/2+hk_j/2T_j(E_0))(T_j(E_0+\hbar\lambda/2)-T_j(E_0-\hbar\lambda/2))/\hbar\right],
\end{eqnarray}
and
\begin{eqnarray}
&&\int_{E_0+hk_j/2T_j(E_0)}^{E_0+\epsilon/2}dE~B_{j,j}(E) \cos
(\frac{S_j(E+\hbar\lambda/2)-S_j(E-\hbar\lambda/2)}{\hbar})\sim\nonumber \\
&&\frac{2\hbar
B_{j,j}(E_0)}{T_j(E_0+\hbar\lambda/2)-T_j(E_0-\hbar\lambda/2)} \cos
(\frac{S_j(E_0+\hbar\lambda/2)-S_j(E_0-\hbar\lambda/2)}{\hbar})\cdot\nonumber
\\
&&\sin \left[
(\epsilon/2-hk_j/2T_j(E_0))(T_j(E_0+\hbar\lambda/2)-T_j(E_0-\hbar\lambda/2))/\hbar\right].
\end{eqnarray}
Combined, these last two terms contribute
\begin{eqnarray}
&&4\epsilon 
B_{j,j}(E_0)\cos
(\frac{S_j(E_0+\hbar\lambda/2)-S_j(E_0-\hbar\lambda/2)}{\hbar})\cdot\nonumber
\\
&&{\rm sinc} [
\epsilon(T_j(E_0+\hbar\lambda/2)-T_j(E_0-\hbar\lambda/2))/\hbar]\cdot
\nonumber \\
&&\cos [2\pi k_j
(T_j(E_0+\lambda\hbar/2)-T_j(E_0-\lambda\hbar/2))/T_j(E_0))].
\label{restr3}
\end{eqnarray}
Therefore, in the limit as $\epsilon\rightarrow \langle\Delta E\rangle$ the total
contribution for case 3 is
\begin{eqnarray}
&&\int_{E_0-\epsilon/2}^{E_0+\epsilon/2}dE~B_{j,j}(E) \cos
(\frac{S_j(E+\hbar\lambda/2)-S_j(E-\hbar\lambda/2)}{\hbar})\nonumber
\\
&&\sim
\frac{hk_jB_{j,j}(E_0)}{T_j(E_0)} \cos
(\frac{S_j(E_0+\hbar\lambda/2)-S_j(E_0-\hbar\lambda/2)}{\hbar})\nonumber \\
&&+4\langle\Delta E\rangle B_{j,j}(E_0)\cos
(\frac{S_j(E_0+\hbar\lambda/2)-S_j(E_0-\hbar\lambda/2)}{\hbar})\cdot
\nonumber \\
&&\cos [2\pi k_j
(T_j(E_0+\lambda\hbar/2)-T_j(E_0-\lambda\hbar/2))/T_j(E_0))].
\label{cont4}
\end{eqnarray}

In summary, evaluation of the three types of integrals in Eq. (\ref{eqyuk}) 
in the correspondence limit, yields the contributions of Eqs. (\ref{cont1}), 
(\ref{cont2}), (\ref{cont3}) and (\ref{cont4}) which combined make a total
contribution of the form 
\begin{equation}
\frac{hk_jB_{j,j}(E_0)}{T_j(E_0)}\cos (\frac{S_j(E_0+\lambda\hbar/2)-S_j(E_0-\hbar\lambda/2)}{\hbar})+O(h^{s-1}e^{iz/\hbar}).
\end{equation}

\pagebreak


\begin{thebibliography}{99}
\bibitem{IVvonN} A quantum system of two or more degrees of freedom is
ergodic if its Hamiltonian alone constitutes the complete set of 
commuting observables, and the energy eigenspectrum is nondegenerate.
\bibitem{IVhyper} We will make the assumption that the classical systems
discussed in this chapter are hyperbolic. This means that
they are ergodic, weak mixing, mixing, etc. \cite{IVarnold}. Furthermore
all periodic orbits for such systems are unstable.
\bibitem{IVaphase} Note that the spectral projection operators, unlike
the Liouville eigenfunctions in both classical and 
quantum mechanics, have a uniquely determined phase.
\bibitem{IVherm} These operators are Hermitian in the Hilbert associated
with the Liouville dynamics, i.e., the space spanned by the vectors
$|n,m)$.
\bibitem{IVobs} Alternatively, the dynamics of observables is governed
by the equation $\frac{\partial}{\partial t}|A(t))=iL|A (t))$, so that
$(\rho (t)|A(0))=(\rho (0)|A(t))$.
\bibitem{IVformula} The identity in Eq. (\ref{subid1p}) is valid only
when the energy spectrum exhibits level repulsion, i.e., is
non-degenerate.
\bibitem{IVsc} For systems with two degrees of freedom $\sigma_j$ is
twice the number of revolutions of the unstable (stable) manifold
about the periodic orbit during one traversal of the
orbit\cite{IVcrl}.
\bibitem{IVconv} Technically, the uniform energy average employed here
will not guarantee correspondence since the resultant damping is
proportional to $1/T_j$ while the periodic orbits
proliferate\cite{IVhannay,IVberryd3} as the
exponential of $T_j$. However, the arguments of Sec. \ref{4.4.2} are
independent of the method of averaging, and can be restated in terms
of a Gaussian exponential average which guarantees convergence\cite{IVberry4}.
We adopt the uniform energy average for simplicity.
\bibitem{footpi} Here $\epsilon\rightarrow \langle
\Delta E\rangle$ whereas is in the eigenfunction section the limit was
$\pi \epsilon\rightarrow \langle \Delta E\rangle$. The difference of $\pi$
can be traced to the different choice of energy smoothing in two sections.
\bibitem{broad} To obtain Eq. (\ref{mugutz}) note that
$d_{\mu}(E)=\sum_n
\Omega_{\mu}(E-E_n)=\int dE'~\Omega_{\mu}(E-E')~d(E')$. By the
convolution theorem this equals $h^{-1}\int
dt~e^{iEt/\hbar}~\tilde{\Omega}_{\mu}(t)~\tilde{d}(t)$, where $\tilde{\Omega}_{\mu}(t)$
and $\tilde{d}(t)$ are the Fourier Transforms of $\Omega_{\mu}(E)$ and
$d(E)$, respectively. This can be written as
$d_{\mu}(E)=h^{-1}\int dt ~\tilde{\Omega}_{\mu}(t)
~{\rm Tr}[e^{i(E-\hat{H})t/\hbar}]$, where $\tilde{\Omega}_{\mu}(t)={\rm sin
c}(\mu t/2\hbar)$. Adopting the semiclassical approximation of Berry
\cite{IVberry3} for the trace in this expression, and performing the time
integral by stationary phase, then gives Eq. (\ref{mugutz}).
\end{thebibliography}
\end{document}